\documentclass[%
 reprint,
 amsmath,amssymb,
 aps,
 prl,
floatfix,
]{revtex4-2}

\usepackage{upgreek}
\usepackage[dvipsnames]{xcolor} 

\usepackage{graphicx}
\usepackage{dcolumn}
\usepackage{bm}

\begin{document}


\title{Injection and Acceleration of Electrons by Radially Polarized Laser Pulses in a Plasma Channel}

\author{P. Hadjisolomou}
\email{Prokopis.Hadjisolomou@eli-beams.eu}
\affiliation{ELI Beamlines Facility, Extreme Light Infrastructure ERIC, Za Radnicí 835, 25241 Dolní Břežany, Czech Republic}

\author{P. Valenta}
\affiliation{ELI Beamlines Facility, Extreme Light Infrastructure ERIC, Za Radnicí 835, 25241 Dolní Břežany, Czech Republic}

\author{R. Shaisultanov}
\affiliation{ELI Beamlines Facility, Extreme Light Infrastructure ERIC, Za Radnicí 835, 25241 Dolní Břežany, Czech Republic}

\author{T. M. Jeong}
\affiliation{ELI Beamlines Facility, Extreme Light Infrastructure ERIC, Za Radnicí 835, 25241 Dolní Břežany, Czech Republic}

\author{D. Gorlova}
\affiliation{York Plasma Institute, Department of Physics, University of York, Heslington, York, North Yorkshire YO10 5DD, UK}

\author{C. P. Ridgers}
\affiliation{York Plasma Institute, Department of Physics, University of York, Heslington, York, North Yorkshire YO10 5DD, UK}

\author{S. V. Bulanov}
\affiliation{ELI Beamlines Facility, Extreme Light Infrastructure ERIC, Za Radnicí 835, 25241 Dolní Břežany, Czech Republic}

\date{\today}

\begin{abstract}
We consider injection and subsequent acceleration of electrons in narrow plasma channels irradiated by linearly and radially polarized ultraintense laser pulses. Using three-dimensional particle-in-cell simulations, we show that radially polarized beams significantly promote electron release from the channel walls and lead to enhanced injection. We compare an $\mathrm{f}/10$ linearly polarized laser beam with two radially polarized cases: one focused more tightly ($\mathrm{f}/5$) to match peak intensity, and one at equal $\mathrm{f}/10$ to capture polarization effects. The radially polarized $\mathrm{f}/10$ case injects approximately one-third more charge than the linearly polarized case, while the $\mathrm{f}/5$ radially polarized case outperforms the linearly polarized one by about a factor of two in terms of maximum electron energy. These results highlight polarization and focusing geometry as key parameters for optimizing laser-driven electron acceleration setups.
\end{abstract}

\maketitle





Direct laser acceleration (DLA) is a mechanism by which relativistic electrons gain energy directly from the oscillating laser field—a concept proposed over two decades ago \cite{1998_PukhovA, 1999_PukhovA, 1999_GahnG}. Compared to the laser wakefield acceleration (LWFA) scheme \cite{1979_TajimaT, 2009_EsareyE, 2016_BulanovSV, 2019_GonsalvesAJ, 2024_PicksleyA, 2024_LazzariniCM}, DLA does not rely on plasma wave excitation and can deliver higher total charge over shorter distances \cite{2021_HusseinAE, 2021_ShawJL}. However, it typically results in broader energy spectra, which has limited its appeal for applications requiring monoenergetic beams. Recent advances in high-power laser systems and beam shaping have intensified interest in DLA \cite{2020_RosmejON, 2024_BabjakR, 2024_BabjakRb, 2024_CohenI, 2024_TangH, 2025_RosmejON}, particularly for scenarios where beam charge outweighs the need for energy resolution.

Traditionally, most DLA studies have employed linearly polarized (LP) lasers in plasma channels, either preformed \cite{2016_StarkDJ, 2018_VranicM, 2020_GongZ, 2020_JirkaM, 2023_MartinezB, 2024_ValentaP} or self-generated \cite{2016_ArefievAV, 2021_LiFY}. In these configurations, motion perpendicular to the polarization plane is typically considered negligible. However, the role of laser polarization in shaping the dynamics of electron injection and acceleration remains relatively unexplored. This is somewhat surprising, as polarization directly determines the spatial structure of the laser’s electric and magnetic fields, and thus the forces acting on electrons near sharp plasma boundaries.

Recent theoretical work emphasized that radial polarization has a strong longitudinal component of the electric field, $E_x$, along the laser propagation direction, assumed here to be the $x$-axis \cite{2018_JeongTM}. In cylindrical waveguides, the symmetry of radially polarized (RP) pulses naturally complements the geometry, potentially enhancing the field overlap with wall-bound electrons and improving injection efficiency. Despite this theoretical promise, there has been a lack of studies of DLA driven by RP pulses, particularly in fully three-dimensional (3D) geometries.

Our work aims to fill this gap by quantitatively analyzing electron injection efficiency and collimation in cylindrical plasma channels irradiated by ultraintense laser pulses with different polarizations. We compare a LP $\mathrm{f}/10$ beam with two RP pulses—one at $\mathrm{f}/5$ to match peak intensity, and one at $\mathrm{f}/10$ to capture polarization effects without modifying the optical setup.


To find the accelerated electron energy scaling, we assume a rectangular waveguide \cite{2011_PozarM}, along the $x$-axis. The $\mathrm{TE_{10}}$ (transverse electric) mode is the lowest one. Electromagnetic modes in the waveguide satisfy the dispersion equation, $\omega^2 = c^2\!\left(k_x^2+\kappa^2\right)$, where $k_x$ is the longitudinal wavenumber, $\kappa$ is the transverse cutoff wavenumber, $\omega$ is the electromagnetic wave angular frequency and $c$ is the speed of light in vacuum. Thus, the group velocity is $v_g={c\,k_x}/{\sqrt{k_x^2+\kappa^2}}$, the phase velocity is $v_{p}={c\sqrt{k_x^2+\kappa^2}}/{k_x}$ and $v_g\,v_{p}=c^2$. This guided-mode description is formally analogous to the accelerating TM-like modes used in RF linear accelerators, where a longitudinal field component on axis enables sustained energy gain \cite{1984_KapchinskiyIM, 1999_HumphriesS}. For the rectangular $\mathrm{TE_{10}}$ mode for a waveguide of width $l_y$, we have $\kappa={\pi}/{l_y}$.

We transform to the reference frame moving along $+x$ at the group velocity of the wave, $v_g$. The wave vector $(\omega/c,\,k_x)$ transforms as $k'_x=\gamma_g\!\left(k_x-\beta_g {\omega}/{c}\right)=0$ (from which we get $\beta_g=c\,k_x/\omega$) and $\omega'=\gamma_g(\omega-\beta_g c\,k_x)$. Substituting $\beta_g$ into $\omega'$ we obtain $\omega'=\gamma_g \,{c^2 \kappa^2}/{\omega}$, and with $\gamma_g=\omega/(c \kappa)=2 l_y /\lambda$ (where $\lambda = 2 \, \pi \, c / \omega$ is the laser wavelength) we have $\omega'=c\,\kappa={\pi c}/{l_y}$.

Numerical integration of equations of relativistic electron motion in the electric and magnetic field for initial conditions, corresponding to the case when the electron in the laboratory frame of reference is at rest, shows that the electron energy is linearly proportional to the field amplitude,
\begin{equation}
a_{\mathrm{W}}=\frac{e B_0\,l_y}{\pi m_e c}  ,
\label{eq:a_SI_E}
\end{equation}
where $B_0$ is the laser magnetic field amplitude, $m_e$ the electron mass, and $e$ the elementary charge. By combining the Lorentz factor, $\gamma'^2 = 1 + {p'^2}/{(m_e c)^2}$, with $p'_\perp\simeq a_{\mathrm{W}}\,m_e c$ and $p'_x=-m_e c\,p_0$ (where $p_0$ is the dimensionless initial electron momentum in the boosted frame of reference) we get
\begin{equation}
\gamma'=\sqrt{1+\left(\frac{p'_\perp}{m_ec}\right)^2+\left(\frac{p'_x}{m_ec}\right)^2}
= \sqrt{1+a_{\mathrm{W}}^2+p_0^2} .
\end{equation}

Energy transforms as $\mathcal{E}_t=\gamma_g\!\left(\mathcal{E}'+\beta_g c\,p'_x\right)$ with $\mathcal{E}'=\gamma' m_e c^2$, giving $\mathcal{E}_t = m_e c^2\, \gamma_g \left(\gamma' - \beta_g p_0\right)$, or
\begin{equation}
\mathcal{E}_t=m_e c^2\,\gamma_g\!\left(\sqrt{1+a_{\mathrm{W}}^2+p_0^2}-\beta_g p_0\right) ,
\end{equation}
where $p_0 = \beta_g/ \sqrt{1-\beta_g^2} = \sqrt{\gamma_g^2-1}$. Subtracting the rest energy, the kinetic energy is
\begin{equation}
\mathcal{E}=m_e c^2\, \left[ \gamma_g\!\left(\sqrt{1+a_{\mathrm{W}}^2+p_0^2}-\beta_g p_0\right) -1 \right] ,
\label{eq:Ee_compact}
\end{equation}
which simplifies to
\begin{equation}
\mathcal{E} = m_e c^2 \, \gamma_g^2 \left[ \sqrt{1+\left( \frac{a_{\mathrm{W}}}{\gamma_g} \right)^2} - 1 \right] .
\end{equation}

The estimates above implicitly assume that electrons interact with the peak transverse field. In practice, misalignment, transverse drift, and dephasing reduce the effective field sampled. We therefore introduce a reduction factor $0<f<1$ and define $a_{\mathrm{W}}^{\mathrm{(eff)}} = f\,a_{\mathrm{W}}$, which yields
\begin{equation}
\mathcal{E} = m_e c^2 \, \gamma_g^2 \left[ \sqrt{1+\left( f \frac{a_{\mathrm{W}}}{\gamma_g} \right)^2} - 1 \right] .
\label{eq:Ee_compact3}
\end{equation}
Using known parameters of $a_{\mathrm{W}} \approx 2000$, $\lambda = 1\,\mu\mathrm{m}$ and ${l_y}/2 = 9\,\mu\mathrm{m}$, for $f=0.04$ we obtain an indicative energy of about $1 \, \mathrm{GeV}$, serving as a reference for the expected energy scale.



\begin{figure}[t]
\centering
\includegraphics[width=1.0\columnwidth]{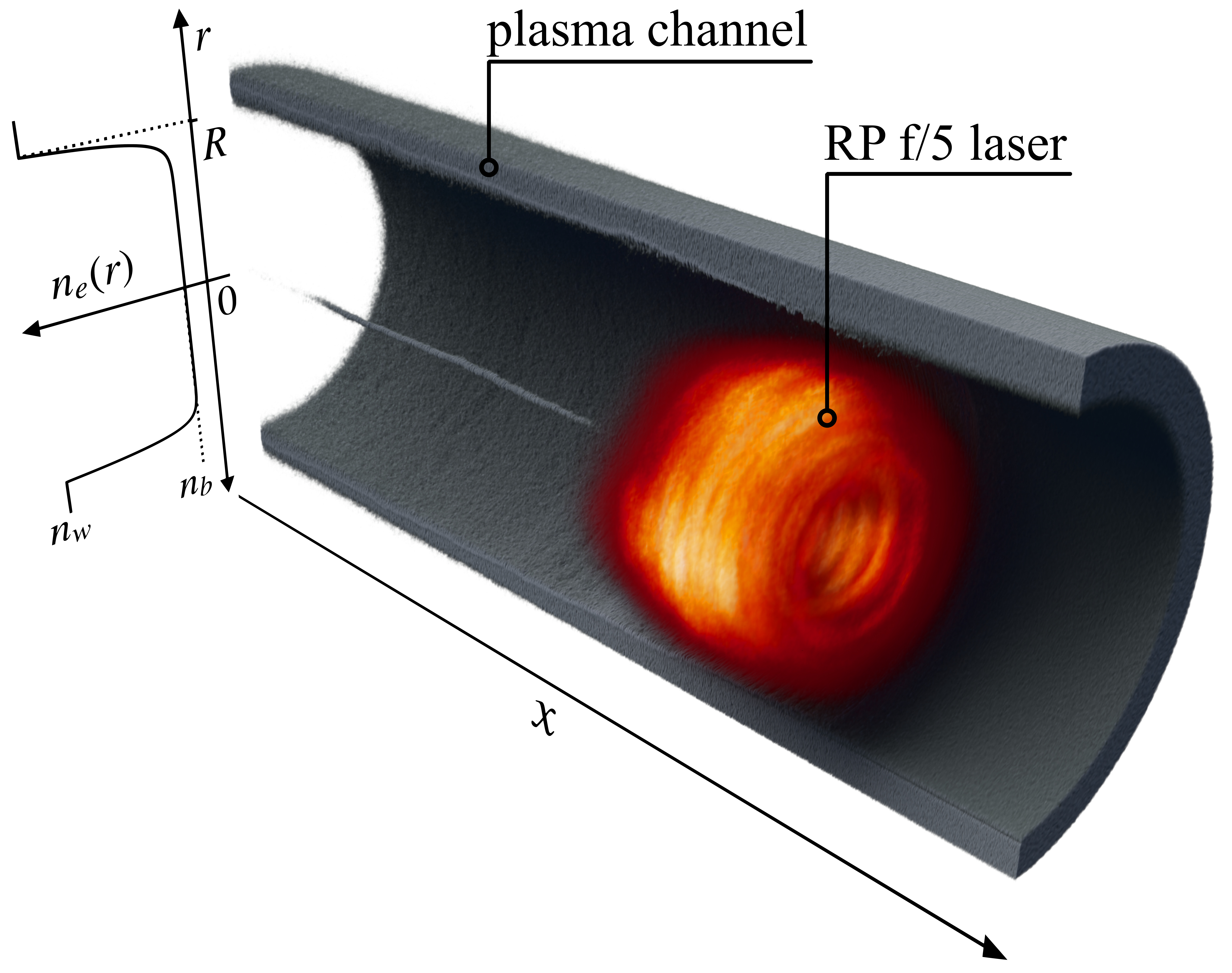}
\caption{Setup of the PIC simulation for the RP $\mathrm{f}/5$ configuration. The laser pulse (orange) propagates inside a preformed narrow plasma channel (gray). The radial electron density of the channel is shown in the figure by $n_{e}(r)$. An aspect ratio of \( 1\!:\!4 \; (\text{x-axis} : \text{r-axis}) \) is applied in the figure.}
\label{fig1}
\end{figure}

We perform fully 3D particle-in-cell (PIC) simulations using the EPOCH code \cite{2015_ArberTD}, shown in Fig.~\ref{fig1}. The laser pulse has $\lambda = 1 \,\mathrm{\upmu m}$, a full-width-at-half-maximum (FWHM) duration of $150 \,\mathrm{fs}$, and a peak intensity of $4.24 \times 10^{21}\,\mathrm{W\,cm^{-2}}$, corresponding to a normalized field amplitude of $a_0 \approx 55$. For the LP case, the beam is focused to a waist of $w_0 = 10\,\mathrm{\upmu m}$. To ensure consistent comparison, we simulate two RP cases: one with tighter focusing at $\mathrm{f}/5$ ($w_0 = 5\,\mathrm{\upmu m}$) to match the LP peak intensity, and another at $\mathrm{f}/10$ with the same waist as the LP configuration to isolate polarization effects. In all cases, the laser energy and power are fixed at approximately $ 1\,\mathrm{kJ}$ and $6.7\,\mathrm{PW}$, respectively.

The plasma consists of a fully ionized channel with a wall electron number density of $n_w = 10 \, n_c$ (where $n_c = \varepsilon_0 \, m_e \, \omega^2 / e^2$ is the critical plasma density, with $\varepsilon_0$ the vacuum permittivity) and an atomic-to-charge ratio of $A/Z = 2.3$. The channel has an inner radius, $R$, of $10\,\mathrm{\upmu m}$ and a wall thickness of $2 \,\mathrm{\upmu m}$, sufficient to prevent wall break-up over the simulation duration of $1.5 \, \mathrm{ps}$. The channel interior is filled with an exponential density ramp, defined for $r < R$ as
\begin{equation}
n_e(r) = n_b + (n_w - n_b)\,\exp\!\left(\frac{r - R}{L_r}\right),
\end{equation}
where $n_b = 10^{-3}\,n_c$ is the on-axis density and $L_r = 0.5\,\mathrm{\upmu m}$ is the exponential scale length (see Figure \ref{fig1}). The front of the channel (excluding the entrance aperture) is sealed with a $5\,\mathrm{\upmu m}$ thick foil to suppress ‘low-amplitude’ fields at the outer surface, which would otherwise reach intensities of $2.4\times10^{20}\,\mathrm{W\,cm^{-2}}$. An equivalent configuration would involve a significantly thicker channel, but such geometry would be computationally prohibitive.

The simulation domain starts at $-92.16 \, \mathrm{\upmu m}$ and ends at $153.6 \, \mathrm{\upmu m}$ in $x$, and spans $\pm 25.75 \, \mathrm{\upmu m}$ in both $y$ and $z$, discretized with $10 \, \mathrm{nm}$ longitudinal and $100 \, \mathrm{nm}$ transverse resolution. The laser is focused at $x = 0$, at the entrance of the channel. Each species is initialized with two macroparticles per cell, totaling over one billion macroparticles per species. A moving window is activated at $742 \, \mathrm{fs}$, propagating in the $+x$ direction at the speed of light.



\begin{figure}[t]
\centering
\includegraphics[width=1.0\columnwidth]{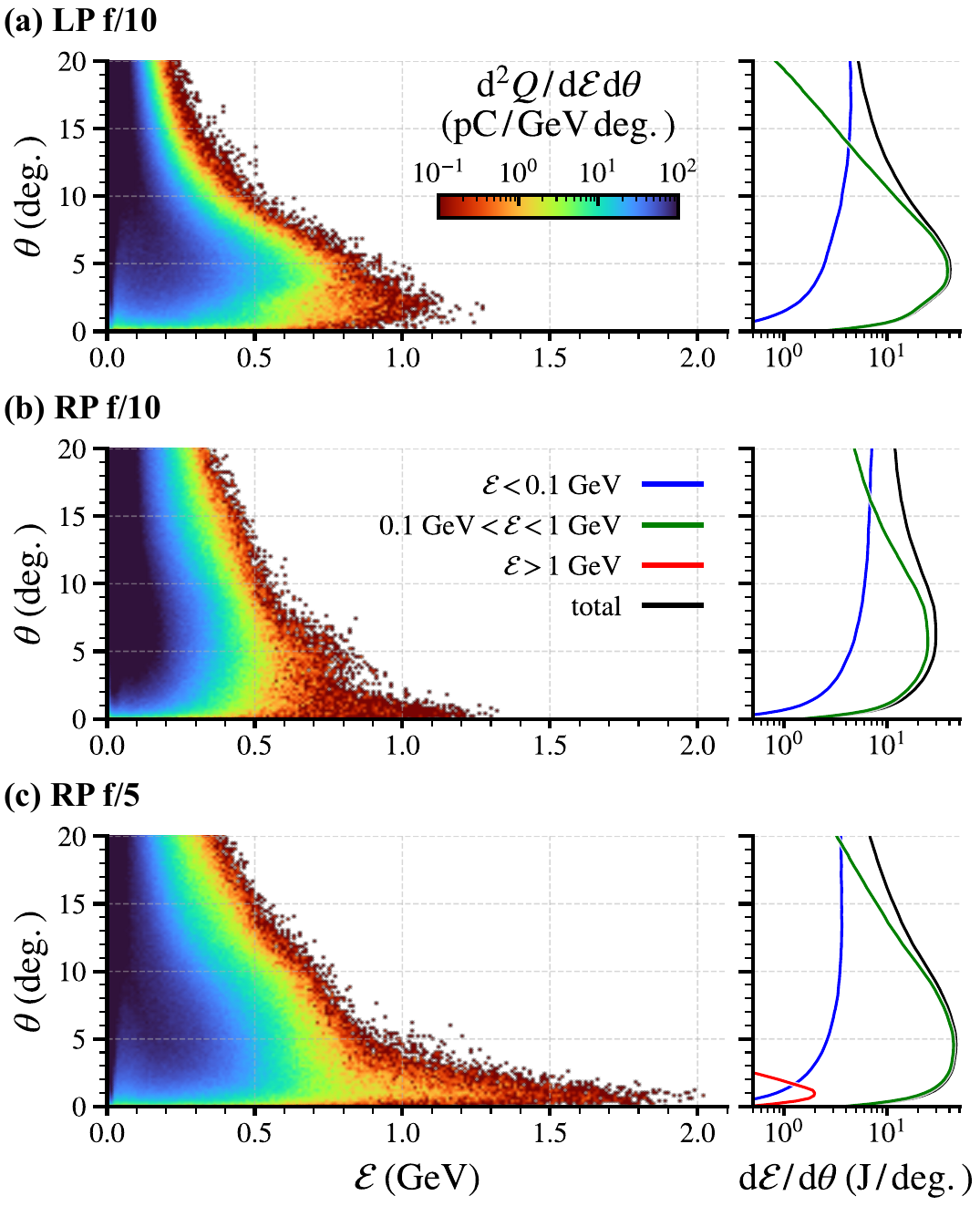}
\caption{Left: Angular and energy spectral distribution of electrons at the end of the simulation, for the three laser configurations: (a) LP $\mathrm{f}/10$, (b) RP $\mathrm{f}/10$, and (c) RP $\mathrm{f}/5$. High-energy electrons exhibit narrower divergence, where for the RP $\mathrm{f}/5$ case a distinct narrow electron population of energy $\gtrapprox 1 \, \mathrm{GeV}$ overlaps to the broader distribution of the moderate-to-low energy electrons. Right: Angular electron energy flux for (a) LP $\mathrm{f}/10$, (b) RP $\mathrm{f}/10$, and (c) RP $\mathrm{f}/5$ (black line). Red, green and blue curves represent energy ranges $\mathcal{E} > 1 \, \mathrm{GeV}$, $0.1 < \mathcal{E}< 1 \, \mathrm{GeV}$, and $\mathcal{E} < 0.1 \, \mathrm{GeV}$, respectively. The $\mathrm{f}/5$ RP case is the only one that forms a peak for electron energies $> 1 \, \mathrm{GeV}$ and a charge threshold $> 0.1 \, \mathrm{nC}$.}
\label{fig2}
\end{figure}

The plasma within the channel has a density of $n_c$ at a radius of approximately $9\,\mathrm{\upmu m}$, which we take as the effective inner radius, $R_e$. In all three simulation cases, the temporal evolution of spatio-angular phase-space plots (Fig.~\ref{fig2}) suggests that electron injection occurs primarily from electrons initially located near the inner surface of the surrounding channel walls \cite{2019_SnyderJ, 2021_WangT}, in accordance with the mechanism described in our previous work \cite{2024_ValentaP}. These electrons are subsequently accelerated via DLA, with a subset reaching kinetic energies in the GeV level.

The left panels of Fig.~\ref{fig2} present the angular and energy spectral distribution of electrons. In all cases, the peak of the (total) angular distribution occurs at $5^\circ$–$7^\circ$. The suppression of particles exactly in the forward direction ($\theta \approx 0^\circ$) suggests the build-up of transverse momentum during acceleration. The forward-axis minima have been observed previously \cite{2017_ShawJL, 2018_ShawJL, 2021_KingPM}. In RP $\mathrm{f}/5$, Fig.~\ref{fig2}(c), the high-energy ($> 1 \, \mathrm{GeV}$) beam is strongly peaked close to the axis, demonstrating high confinement.

In the LP case, the resulting electron beam exhibits a broad angular distribution, characterized by a divergence angle (energy flux at FWHM) of approximately $7^\circ$ and an exponentially decreasing energy spectrum. The RP $\mathrm{f}/10$ configuration, by contrast, yields a higher injected charge in the effective core (approximately 35\% more) than the LP case, but exhibits noticeably poorer collimation. Conversely, the RP $\mathrm{f}/5$ configuration maintains a similar injected charge and angular spread to the LP case, with significantly improved collimation when considering only high-energy electrons ($\mathcal{E} > 1 \,\mathrm{GeV}$). The narrower angular distribution observed in Fig.~\ref{fig2} originates from an acceleration scheme driven by the longitudinal electric field of the radially polarized beam \cite{2006_SalaminYI, 2011_SinghKP, 2016_CarbajoS}. In this regime, electrons are accelerated on axis by the $E_x$ component, forming a confined high-energy bunch that coexists with the broader DLA population. The axially accelerated electron population, although much smaller in number, reaches energies approximately twice those of the pure DLA acceleration scheme when comparing cases of equal peak intensity (RP $\mathrm{f}/5$ vs LP $\mathrm{f}/10$).

Expanding on this, the right panels of Fig.~\ref{fig2} reveal how angular flux patterns vary across energy bands. The RP $\mathrm{f}/10$ configuration dominates in the regime of $\mathcal{E}<0.1 \, \mathrm{GeV}$. The LP $\mathrm{f}/10$ and RP $\mathrm{f}/5$ configurations result in a similar flux profile in the intermediate energy region, peaking at $5^\circ$. Notably, the RP $\mathrm{f}/5$ configuration for $\mathcal{E}>1 \, \mathrm{GeV}$ flux peaks at $1^\circ$ with a divergence of $1.5^\circ$. The charge contained in the high-energy population is $2.4 \, \mathrm{nC}$. For the other two cases the charge is below $0.1 \, \mathrm{nC}$ (for  LP $f/10$ and RP $f/10$ the charge is $0.09 \, \mathrm{nC}$ and $0.03 \, \mathrm{nC}$, respectively) and reliable formation of the high-energy population is not seen.

In all cases, the electron population with energies below $0.1 \, \mathrm{GeV}$ displays a markedly different angular pattern, showing a broad flux distribution between approximately $3^\circ$ and $55^\circ$. This distribution suggests that the low-energy electron population originate from a different spatial region than the collimated, high-energy core. Comparison of 2D field and particle slices indicates that high-energy electrons are predominantly accelerated in regions of strong laser field overlap, while lower-energy electrons originate from the rear of the channel, where a central electron filament (see Figure \ref{fig1}) forms alongside quasi-static magnetic fields of $\mathcal{O}(0.1~\mathrm{MT})$.

\begin{figure}[t]
\centering
\includegraphics[width=0.9\columnwidth]{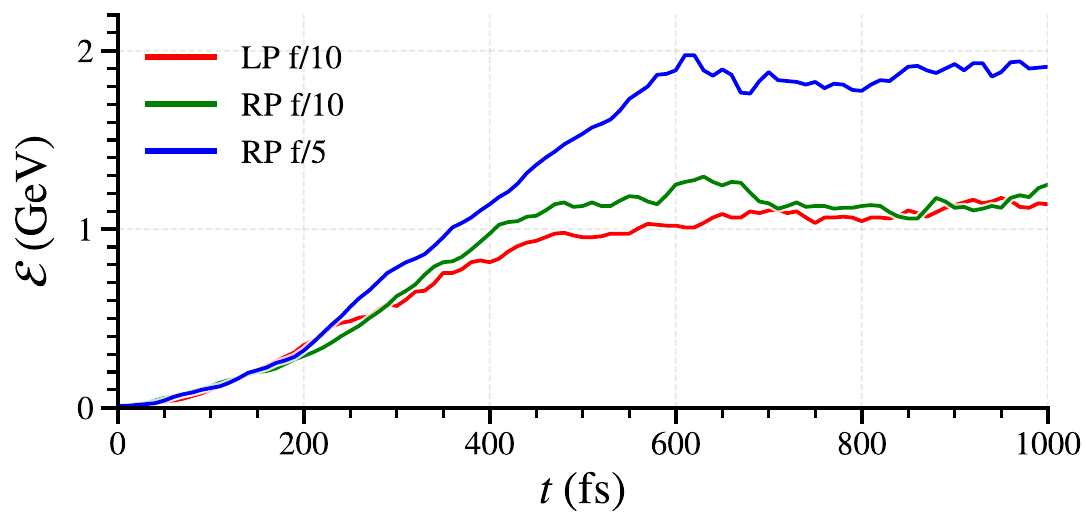}
\caption{Temporal evolution of maximum electron energy for the three laser configurations. RP $\mathrm{f}/5$ continues accelerating electrons up to $600 \, \mathrm{fs}$, reaching nearly $2 \, \mathrm{GeV}$, while both the LP $\mathrm{f}/10$ and RP $\mathrm{f}/10$ saturate near $1.2 \, \mathrm{GeV}$.}
\label{fig3}
\end{figure}

Figure~\ref{fig3} shows the temporal evolution of the maximum electron energy. In the initial phase of the interaction, both the LP $\mathrm{f}/10$ and RP $\mathrm{f}/5$ configurations—which share the same peak intensity—exhibit a similar rate of energy increase, indicating that the early energy gain is distributed across the bulk DLA population (up to ${\sim} \, 1 \,\mathrm{GeV}$). Beyond this energy the curves diverge. The RP $\mathrm{f}/5$ case continues accelerating electrons up to $1.9 \,\mathrm{GeV}$, while both the LP $\mathrm{f}/10$ and RP $\mathrm{f}/10$ configurations saturate near $1.2 \,\mathrm{GeV}$ by $500 \,\mathrm{fs}$. This saturation likely results from a combination of electron channel ejection \cite{2024_ValentaP}, dephasing and reduced overlap with the laser fields. In contrast, the continued energy growth in the RP $\mathrm{f}/5$ case arises from the longitudinal electric field component, which keeps electrons phase-locked with the accelerating (negative-$E_x$) half-cycle for about $50\,\mathrm{fs}$, enabling sustained on-axis acceleration. Although RP $\mathrm{f}/5$ and LP $\mathrm{f}/10$ share identical laser power and intensity, the additional longitudinal-field contribution allows RP $\mathrm{f}/5$ to achieve substantially higher maximum electron energies.

\begin{figure}[t]
\centering
\includegraphics[width=1.0\columnwidth]{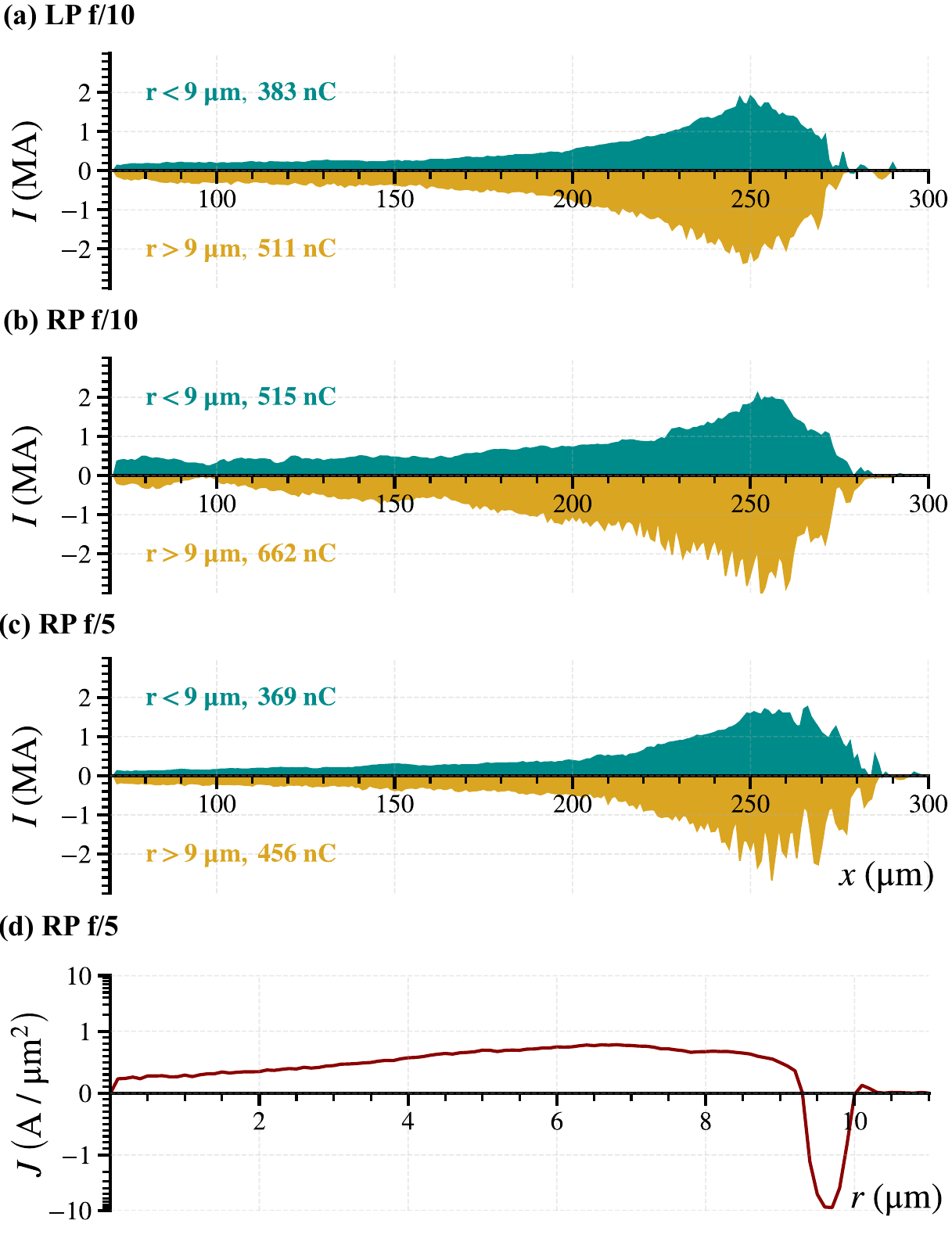}
\caption{Longitudinal profile of current for core electrons ($r < 9 \,\mathrm{\upmu m}$, cyan) and sheath electrons ($r > 9 \,\mathrm{\upmu m}$, mustard) for the (a) LP $\mathrm{f}/10$, (b) RP $\mathrm{f}/10$, and (c) RP $\mathrm{f}/5$ case. RP $\mathrm{f}/10$ carries the largest total charge in both the core and sheath regions, and spreads across an extended tail. Integrated charges for each region are indicated in the legend. (d) Radial profile of the current density for the RP $\mathrm{f}/5$ case, for electrons with $250 \, \mathrm{\upmu m} < x < 260 \, \mathrm{\upmu m}$ (peak current).}
\label{fig4}
\end{figure}

Figure~\ref{fig4} reveals how the accelerated charge is partitioned between the core beam and the surrounding sheath. The positive return current is confined to a narrow annulus extending approximately from $R_e$ to $R$ (sheath thickness of ${\sim} \, 1 \, \mathrm{\upmu m}$, see Figure \ref{fig4}(d)), while the negative current flows in the inner core. All three laser configurations generate a similarly sharp current peak, followed by an extended tail. RP $\mathrm{f}/10$ exhibits a longer tail and accumulates a higher total charge than the LP case, consistent with sustained wall-based injection driven by its radial field components.

In a cylindrical plasma channel where a high-power current is driven through the core, electrons form a strong forward current along the axis, $I_c = e \, n_{el} \, \beta \, c \, \pi \, R_e^2$, where $n_{el}$ is the current electron density and $\beta$ is the normalized velocity. This axial current generates an increasing azimuthal magnetic field around it. According to Lenz's law, any change in magnetic flux induces a current that opposes the change. As the central current strengthens and the magnetic field intensifies, the surrounding plasma responds by forming a return current along the outer regions of the cylinder. In this way, the system self-organizes to oppose the magnetic change induced by the driving current.

The return current partially neutralizes the core’s self-fields and maintains the total current below the Alfvén limit \cite{1939_AlfvenH},
\begin{equation}
I_A = \gamma \, \beta \frac{e \, c}{r_e} \approx 17 \gamma \,\mathrm{kA} ,
\end{equation}
where $\gamma$ is the electron Lorentz factor in the laboratory frame and $r_e = e^2/(4 \, \pi \, \varepsilon_0 \, m_e \, c^2)$ is the classical electron radius. For $a_0 \gg 1$ we have $\gamma \approx a_0$ and the ratio of $I_A$ to $I_c$ gives
\begin{equation}
P_A \approx \frac{a_0}{\pi \, r_e \, n_{el} \, R_e^2} \ge 1
\end{equation}
In the region of peak current, electrons typically have energies of $0.1\,\mathrm{GeV}\text{--}0.4\,\mathrm{GeV}$, corresponding to $\gamma \approx 200\text{--}800$. This yields an Alfvén current threshold of approximately $3.3\,\mathrm{MA}$ for $0.1\,\mathrm{GeV}$ electrons, indicating that the system approaches this limit. By counterbalancing the core current, the sheath helps suppress space-charge divergence and enables the propagation of highly relativistic, dense electron beams without exceeding stability thresholds. If we consider the self-focusing case, in the limit of $P_A=1$ we have
\begin{equation}
R_{min} = \sqrt{\frac{a_0}{\pi \, r_e \, n_{el}}} ,
\end{equation}
which gives the radius of the self-focusing channel.

The peak core current in the RP $\mathrm{f}/10$ case reaches approximately $2.5\,\mathrm{MA}$, about $10\%$ higher than in either the LP $\mathrm{f}/10$ or RP $\mathrm{f}/5$ configurations. This value rivals current levels typical of pulsed-power accelerators, such as diode-driven Z-pinches, which deliver multi-MA beams but require vastly larger and less compact setups \cite{2020_SinarsDB}. In contrast, laser-driven systems typically achieve peak currents in the $10\text{--}100\,\mathrm{kA}$ range, albeit in ultrashort ($< 100\,\mathrm{fs}$) bunches. The multi-MA current observed here therefore arises not from the high-energy population (${\sim} \, 1 \, \mathrm{nC}$ for $\mathcal{E}>1 \,\mathrm{GeV}$) but from the much larger number of lower-energy electrons. This enhanced injection is due to the RP field geometry, which allows pulling electrons inward from the entire inner circumference of the channel in this compact, laser-driven plasma accelerator, producing a denser and more symmetric electron flow toward the axis and sustaining a higher net current.



In conclusion, we have demonstrated through 3D PIC simulations that RP laser pulses significantly enhance electron injection from plasma channel walls compared to LP beams. Radial polarization, particularly under tighter focusing, improves beam collimation, increases peak electron energies, and boosts high-energy flux while maintaining substantial current from side-injected electrons. The $\gamma$-flash \cite{2012_NakamuraT, 2012_RidgersCP} properties are a direct consequence of the electron beam quality. Unveiling the electron motion under various field configurations \cite{2021_HadjisolomouP, 2022_HadjisolomouP, 2025_HadjisolomouP} is required to obtain $\gamma$-rays of controlled properties. These results strongly support the conclusion that RP lasers in narrow plasma channels are more efficient for producing energetic, collimated electron beams, offering promising advantages for advanced electron accelerator and $\gamma$-ray source applications \cite{2023_HadjisolomouP}.

Our findings also indicate that field symmetry alone is insufficient to ensure efficient electron acceleration. Without adequate intensity, RP beams may enable broad injection but fail to sustain energy gain. Instead, the combined influence of polarization, focusing geometry, and field strength determines both injection efficiency and final beam quality.

These results suggest clear experimental pathways for validation. Cylindrical microchannel plate targets can be tested at multi-petawatt laser facilities, and radially polarized modes are accessible with existing laser systems. Experimental measurements of the accelerated electron beam characteristics can directly test the predictions presented here.

More broadly, polarization tailoring emerges as a powerful design lever for optimizing laser acceleration. Facilities such as ELI Beamlines, Apollon, and Vulcan 20-20 could leverage this strategy to enhance beam brightness without increasing laser energy, enabling more efficient and versatile next-generation radiation sources.


\begin{acknowledgments}
This work was supported by the project ``e-INFRA CZ'' (ID:90254) from the Ministry of Education, Youth and Sports of the Czech Republic. The EPOCH code used in this work was in part funded by the UK EPSRC grants EP/G054950/1, EP/G056803/1, EP/G055165/1, EP/M022463/1, and EP/P02212X/1.
\end{acknowledgments}




\bibliography{biblio_2025PRL}

\begin{thebibliography}{46}%
\makeatletter
\providecommand \@ifxundefined [1]{%
 \@ifx{#1\undefined}
}%
\providecommand \@ifnum [1]{%
 \ifnum #1\expandafter \@firstoftwo
 \else \expandafter \@secondoftwo
 \fi
}%
\providecommand \@ifx [1]{%
 \ifx #1\expandafter \@firstoftwo
 \else \expandafter \@secondoftwo
 \fi
}%
\providecommand \natexlab [1]{#1}%
\providecommand \enquote  [1]{``#1''}%
\providecommand \bibnamefont  [1]{#1}%
\providecommand \bibfnamefont [1]{#1}%
\providecommand \citenamefont [1]{#1}%
\providecommand \href@noop [0]{\@secondoftwo}%
\providecommand \href [0]{\begingroup \@sanitize@url \@href}%
\providecommand \@href[1]{\@@startlink{#1}\@@href}%
\providecommand \@@href[1]{\endgroup#1\@@endlink}%
\providecommand \@sanitize@url [0]{\catcode `\\12\catcode `\$12\catcode
  `\&12\catcode `\#12\catcode `\^12\catcode `\_12\catcode `\%12\relax}%
\providecommand \@@startlink[1]{}%
\providecommand \@@endlink[0]{}%
\providecommand \url  [0]{\begingroup\@sanitize@url \@url }%
\providecommand \@url [1]{\endgroup\@href {#1}{\urlprefix }}%
\providecommand \urlprefix  [0]{URL }%
\providecommand \Eprint [0]{\href }%
\providecommand \doibase [0]{https://doi.org/}%
\providecommand \selectlanguage [0]{\@gobble}%
\providecommand \bibinfo  [0]{\@secondoftwo}%
\providecommand \bibfield  [0]{\@secondoftwo}%
\providecommand \translation [1]{[#1]}%
\providecommand \BibitemOpen [0]{}%
\providecommand \bibitemStop [0]{}%
\providecommand \bibitemNoStop [0]{.\EOS\space}%
\providecommand \EOS [0]{\spacefactor3000\relax}%
\providecommand \BibitemShut  [1]{\csname bibitem#1\endcsname}%
\let\auto@bib@innerbib\@empty
\bibitem [{\citenamefont {Pukhov}\ and\ \citenamefont {Meyer-ter
  Vehn}(1998)}]{1998_PukhovA}%
  \BibitemOpen
  \bibfield  {author} {\bibinfo {author} {\bibfnamefont {A.}~\bibnamefont
  {Pukhov}}\ and\ \bibinfo {author} {\bibfnamefont {J.}~\bibnamefont {Meyer-ter
  Vehn}},\ }\bibfield  {title} {\bibinfo {title} {{Relativistic laser-plasma
  interaction by multi-dimensional particle-in-cell simulations}},\ }\href
  {https://doi.org/10.1063/1.872821} {\bibfield  {journal} {\bibinfo  {journal}
  {Phys. Plasmas}\ }\textbf {\bibinfo {volume} {5}},\ \bibinfo {pages} {1880}
  (\bibinfo {year} {1998})}\BibitemShut {NoStop}%
\bibitem [{\citenamefont {Pukhov}\ \emph {et~al.}(1999)\citenamefont {Pukhov},
  \citenamefont {Sheng},\ and\ \citenamefont {Meyer-ter Vehn}}]{1999_PukhovA}%
  \BibitemOpen
  \bibfield  {author} {\bibinfo {author} {\bibfnamefont {A.}~\bibnamefont
  {Pukhov}}, \bibinfo {author} {\bibfnamefont {Z.~M.}\ \bibnamefont {Sheng}},\
  and\ \bibinfo {author} {\bibfnamefont {J.}~\bibnamefont {Meyer-ter Vehn}},\
  }\bibfield  {title} {\bibinfo {title} {{Particle acceleration in relativistic
  laser channels}},\ }\href {https://doi.org/10.1063/1.873242} {\bibfield
  {journal} {\bibinfo  {journal} {Phys. Plasmas}\ }\textbf {\bibinfo {volume}
  {6}},\ \bibinfo {pages} {2847} (\bibinfo {year} {1999})}\BibitemShut
  {NoStop}%
\bibitem [{\citenamefont {Gahn}\ \emph {et~al.}(1999)\citenamefont {Gahn},
  \citenamefont {Tsakiris}, \citenamefont {Pukhov}, \citenamefont {Meyer-ter
  Vehn}, \citenamefont {Pretzler}, \citenamefont {Thirolf}, \citenamefont
  {Habs},\ and\ \citenamefont {Witte}}]{1999_GahnG}%
  \BibitemOpen
  \bibfield  {author} {\bibinfo {author} {\bibfnamefont {C.}~\bibnamefont
  {Gahn}}, \bibinfo {author} {\bibfnamefont {G.~D.}\ \bibnamefont {Tsakiris}},
  \bibinfo {author} {\bibfnamefont {A.}~\bibnamefont {Pukhov}}, \bibinfo
  {author} {\bibfnamefont {J.}~\bibnamefont {Meyer-ter Vehn}}, \bibinfo
  {author} {\bibfnamefont {G.}~\bibnamefont {Pretzler}}, \bibinfo {author}
  {\bibfnamefont {P.}~\bibnamefont {Thirolf}}, \bibinfo {author} {\bibfnamefont
  {D.}~\bibnamefont {Habs}},\ and\ \bibinfo {author} {\bibfnamefont {K.~J.}\
  \bibnamefont {Witte}},\ }\bibfield  {title} {\bibinfo {title} {Multi-mev
  electron beam generation by direct laser acceleration in high-density plasma
  channels},\ }\href {https://doi.org/10.1103/PhysRevLett.83.4772} {\bibfield
  {journal} {\bibinfo  {journal} {Phys. Rev. Lett.}\ }\textbf {\bibinfo
  {volume} {83}},\ \bibinfo {pages} {4772} (\bibinfo {year}
  {1999})}\BibitemShut {NoStop}%
\bibitem [{\citenamefont {Tajima}\ and\ \citenamefont
  {Dawson}(1979)}]{1979_TajimaT}%
  \BibitemOpen
  \bibfield  {author} {\bibinfo {author} {\bibfnamefont {T.}~\bibnamefont
  {Tajima}}\ and\ \bibinfo {author} {\bibfnamefont {J.~M.}\ \bibnamefont
  {Dawson}},\ }\bibfield  {title} {\bibinfo {title} {{Laser Electron
  Accelerator}},\ }\href {https://doi.org/10.1103/PhysRevLett.43.267}
  {\bibfield  {journal} {\bibinfo  {journal} {Phys. Rev. Lett.}\ }\textbf
  {\bibinfo {volume} {43}},\ \bibinfo {pages} {267} (\bibinfo {year}
  {1979})}\BibitemShut {NoStop}%
\bibitem [{\citenamefont {Esarey}\ \emph {et~al.}(2009)\citenamefont {Esarey},
  \citenamefont {Schroeder},\ and\ \citenamefont {Leemans}}]{2009_EsareyE}%
  \BibitemOpen
  \bibfield  {author} {\bibinfo {author} {\bibfnamefont {E.}~\bibnamefont
  {Esarey}}, \bibinfo {author} {\bibfnamefont {C.~B.}\ \bibnamefont
  {Schroeder}},\ and\ \bibinfo {author} {\bibfnamefont {W.~P.}\ \bibnamefont
  {Leemans}},\ }\bibfield  {title} {\bibinfo {title} {{Physics of laser-driven
  plasma-based electron accelerators}},\ }\href
  {https://doi.org/10.1103/RevModPhys.81.1229} {\bibfield  {journal} {\bibinfo
  {journal} {Rev. Mod. Phys.}\ }\textbf {\bibinfo {volume} {81}},\ \bibinfo
  {pages} {1229} (\bibinfo {year} {2009})}\BibitemShut {NoStop}%
\bibitem [{\citenamefont {Bulanov}\ \emph {et~al.}(2016)\citenamefont
  {Bulanov}, \citenamefont {Esirkepov}, \citenamefont {Hayashi}, \citenamefont
  {Kiriyama}, \citenamefont {Koga}, \citenamefont {Kotaki}, \citenamefont
  {Mori},\ and\ \citenamefont {Kando}}]{2016_BulanovSV}%
  \BibitemOpen
  \bibfield  {author} {\bibinfo {author} {\bibfnamefont {S.~V.}\ \bibnamefont
  {Bulanov}}, \bibinfo {author} {\bibfnamefont {T.~Z.}\ \bibnamefont
  {Esirkepov}}, \bibinfo {author} {\bibfnamefont {Y.}~\bibnamefont {Hayashi}},
  \bibinfo {author} {\bibfnamefont {H.}~\bibnamefont {Kiriyama}}, \bibinfo
  {author} {\bibfnamefont {J.~K.}\ \bibnamefont {Koga}}, \bibinfo {author}
  {\bibfnamefont {H.}~\bibnamefont {Kotaki}}, \bibinfo {author} {\bibfnamefont
  {M.}~\bibnamefont {Mori}},\ and\ \bibinfo {author} {\bibfnamefont
  {M.}~\bibnamefont {Kando}},\ }\bibfield  {title} {\bibinfo {title} {{On some
  theoretical problems of laser wake-field accelerators}},\ }\href
  {https://doi.org/10.1017/S0022377816000623} {\bibfield  {journal} {\bibinfo
  {journal} {J. Plasma Phys.}\ }\textbf {\bibinfo {volume} {82}},\ \bibinfo
  {pages} {905820308} (\bibinfo {year} {2016})}\BibitemShut {NoStop}%
\bibitem [{\citenamefont {Gonsalves}\ \emph {et~al.}(2019)\citenamefont
  {Gonsalves}, \citenamefont {Nakamura}, \citenamefont {Daniels}, \citenamefont
  {Benedetti}, \citenamefont {Pieronek}, \citenamefont {de~Raadt},
  \citenamefont {Steinke}, \citenamefont {Bin}, \citenamefont {Bulanov},
  \citenamefont {van Tilborg} \emph {et~al.}}]{2019_GonsalvesAJ}%
  \BibitemOpen
  \bibfield  {author} {\bibinfo {author} {\bibfnamefont {A.~J.}\ \bibnamefont
  {Gonsalves}}, \bibinfo {author} {\bibfnamefont {K.}~\bibnamefont {Nakamura}},
  \bibinfo {author} {\bibfnamefont {J.}~\bibnamefont {Daniels}}, \bibinfo
  {author} {\bibfnamefont {C.}~\bibnamefont {Benedetti}}, \bibinfo {author}
  {\bibfnamefont {C.}~\bibnamefont {Pieronek}}, \bibinfo {author}
  {\bibfnamefont {T.~C.~H.}\ \bibnamefont {de~Raadt}}, \bibinfo {author}
  {\bibfnamefont {S.}~\bibnamefont {Steinke}}, \bibinfo {author} {\bibfnamefont
  {J.~H.}\ \bibnamefont {Bin}}, \bibinfo {author} {\bibfnamefont {S.~S.}\
  \bibnamefont {Bulanov}}, \bibinfo {author} {\bibfnamefont {J.}~\bibnamefont
  {van Tilborg}}, \emph {et~al.},\ }\bibfield  {title} {\bibinfo {title}
  {Petawatt laser guiding and electron beam acceleration to 8 gev in a
  laser-heated capillary discharge waveguide},\ }\href
  {https://doi.org/10.1103/PhysRevLett.122.084801} {\bibfield  {journal}
  {\bibinfo  {journal} {Phys. Rev. Lett.}\ }\textbf {\bibinfo {volume} {122}},\
  \bibinfo {pages} {084801} (\bibinfo {year} {2019})}\BibitemShut {NoStop}%
\bibitem [{\citenamefont {Picksley}\ \emph {et~al.}(2024)\citenamefont
  {Picksley}, \citenamefont {Stackhouse}, \citenamefont {Benedetti},
  \citenamefont {Nakamura}, \citenamefont {Tsai}, \citenamefont {Li},
  \citenamefont {Miao}, \citenamefont {Shrock}, \citenamefont {Rockafellow},
  \citenamefont {Milchberg} \emph {et~al.}}]{2024_PicksleyA}%
  \BibitemOpen
  \bibfield  {author} {\bibinfo {author} {\bibfnamefont {A.}~\bibnamefont
  {Picksley}}, \bibinfo {author} {\bibfnamefont {J.}~\bibnamefont
  {Stackhouse}}, \bibinfo {author} {\bibfnamefont {C.}~\bibnamefont
  {Benedetti}}, \bibinfo {author} {\bibfnamefont {K.}~\bibnamefont {Nakamura}},
  \bibinfo {author} {\bibfnamefont {H.~E.}\ \bibnamefont {Tsai}}, \bibinfo
  {author} {\bibfnamefont {R.}~\bibnamefont {Li}}, \bibinfo {author}
  {\bibfnamefont {B.}~\bibnamefont {Miao}}, \bibinfo {author} {\bibfnamefont
  {J.~E.}\ \bibnamefont {Shrock}}, \bibinfo {author} {\bibfnamefont
  {E.}~\bibnamefont {Rockafellow}}, \bibinfo {author} {\bibfnamefont {H.~M.}\
  \bibnamefont {Milchberg}}, \emph {et~al.},\ }\bibfield  {title} {\bibinfo
  {title} {{Matched Guiding and Controlled Injection in Dark-Current-Free,
  10-GeV-Class, Channel-Guided Laser-Plasma Accelerators}},\ }\href
  {https://doi.org/10.1103/PhysRevLett.133.255001} {\bibfield  {journal}
  {\bibinfo  {journal} {Phys. Rev. Lett.}\ }\textbf {\bibinfo {volume} {133}},\
  \bibinfo {pages} {255001} (\bibinfo {year} {2024})}\BibitemShut {NoStop}%
\bibitem [{\citenamefont {Lazzarini}\ \emph {et~al.}(2024)\citenamefont
  {Lazzarini}, \citenamefont {Grittani}, \citenamefont {Valenta}, \citenamefont
  {Zymak}, \citenamefont {Antipenkov}, \citenamefont {Chaulagain},
  \citenamefont {Goncalves}, \citenamefont {Grenfell}, \citenamefont {Lamač},
  \citenamefont {Lorenz}, \citenamefont {Nevrkla}, \citenamefont {Špaček},
  \citenamefont {Šobr}, \citenamefont {Szuba}, \citenamefont {Bakule},
  \citenamefont {Korn},\ and\ \citenamefont {Bulanov}}]{2024_LazzariniCM}%
  \BibitemOpen
  \bibfield  {author} {\bibinfo {author} {\bibfnamefont {C.~M.}\ \bibnamefont
  {Lazzarini}}, \bibinfo {author} {\bibfnamefont {G.~M.}\ \bibnamefont
  {Grittani}}, \bibinfo {author} {\bibfnamefont {P.}~\bibnamefont {Valenta}},
  \bibinfo {author} {\bibfnamefont {I.}~\bibnamefont {Zymak}}, \bibinfo
  {author} {\bibfnamefont {R.}~\bibnamefont {Antipenkov}}, \bibinfo {author}
  {\bibfnamefont {U.}~\bibnamefont {Chaulagain}}, \bibinfo {author}
  {\bibfnamefont {L.~V.~N.}\ \bibnamefont {Goncalves}}, \bibinfo {author}
  {\bibfnamefont {A.}~\bibnamefont {Grenfell}}, \bibinfo {author}
  {\bibfnamefont {M.}~\bibnamefont {Lamač}}, \bibinfo {author} {\bibfnamefont
  {S.}~\bibnamefont {Lorenz}}, \bibinfo {author} {\bibfnamefont
  {M.}~\bibnamefont {Nevrkla}}, \bibinfo {author} {\bibfnamefont
  {A.}~\bibnamefont {Špaček}}, \bibinfo {author} {\bibfnamefont
  {V.}~\bibnamefont {Šobr}}, \bibinfo {author} {\bibfnamefont
  {W.}~\bibnamefont {Szuba}}, \bibinfo {author} {\bibfnamefont
  {P.}~\bibnamefont {Bakule}}, \bibinfo {author} {\bibfnamefont
  {G.}~\bibnamefont {Korn}},\ and\ \bibinfo {author} {\bibfnamefont {S.~V.}\
  \bibnamefont {Bulanov}},\ }\bibfield  {title} {\bibinfo {title}
  {{Ultrarelativistic electron beams accelerated by terawatt scalable kHz
  laser}},\ }\href {https://doi.org/10.1063/5.0189051} {\bibfield  {journal}
  {\bibinfo  {journal} {Phys. Plasmas}\ }\textbf {\bibinfo {volume} {31}},\
  \bibinfo {pages} {030703} (\bibinfo {year} {2024})}\BibitemShut {NoStop}%
\bibitem [{\citenamefont {Hussein}\ \emph {et~al.}(2021)\citenamefont
  {Hussein}, \citenamefont {Arefiev}, \citenamefont {Batson}, \citenamefont
  {Chen}, \citenamefont {Craxton}, \citenamefont {Davies}, \citenamefont
  {Froula}, \citenamefont {Gong}, \citenamefont {Haberberger}, \citenamefont
  {Ma} \emph {et~al.}}]{2021_HusseinAE}%
  \BibitemOpen
  \bibfield  {author} {\bibinfo {author} {\bibfnamefont {A.~E.}\ \bibnamefont
  {Hussein}}, \bibinfo {author} {\bibfnamefont {A.~V.}\ \bibnamefont
  {Arefiev}}, \bibinfo {author} {\bibfnamefont {T.}~\bibnamefont {Batson}},
  \bibinfo {author} {\bibfnamefont {H.}~\bibnamefont {Chen}}, \bibinfo {author}
  {\bibfnamefont {R.~S.}\ \bibnamefont {Craxton}}, \bibinfo {author}
  {\bibfnamefont {A.~S.}\ \bibnamefont {Davies}}, \bibinfo {author}
  {\bibfnamefont {D.~H.}\ \bibnamefont {Froula}}, \bibinfo {author}
  {\bibfnamefont {Z.}~\bibnamefont {Gong}}, \bibinfo {author} {\bibfnamefont
  {D.}~\bibnamefont {Haberberger}}, \bibinfo {author} {\bibfnamefont
  {Y.}~\bibnamefont {Ma}}, \emph {et~al.},\ }\bibfield  {title} {\bibinfo
  {title} {Towards the optimisation of direct laser acceleration},\ }\href
  {https://doi.org/10.1088/1367-2630/abdf9a} {\bibfield  {journal} {\bibinfo
  {journal} {New J. Phys.}\ }\textbf {\bibinfo {volume} {23}},\ \bibinfo
  {pages} {023031} (\bibinfo {year} {2021})}\BibitemShut {NoStop}%
\bibitem [{\citenamefont {Shaw}\ \emph {et~al.}(2021)\citenamefont {Shaw},
  \citenamefont {Romo-Gonzalez}, \citenamefont {Lemos}, \citenamefont {King},
  \citenamefont {Bruhaug}, \citenamefont {Miller}, \citenamefont {Dorrer},
  \citenamefont {Kruschwitz}, \citenamefont {Waxer}, \citenamefont {Williams}
  \emph {et~al.}}]{2021_ShawJL}%
  \BibitemOpen
  \bibfield  {author} {\bibinfo {author} {\bibfnamefont {J.~L.}\ \bibnamefont
  {Shaw}}, \bibinfo {author} {\bibfnamefont {M.~A.}\ \bibnamefont
  {Romo-Gonzalez}}, \bibinfo {author} {\bibfnamefont {N.}~\bibnamefont
  {Lemos}}, \bibinfo {author} {\bibfnamefont {P.~M.}\ \bibnamefont {King}},
  \bibinfo {author} {\bibfnamefont {G.}~\bibnamefont {Bruhaug}}, \bibinfo
  {author} {\bibfnamefont {K.~G.}\ \bibnamefont {Miller}}, \bibinfo {author}
  {\bibfnamefont {C.}~\bibnamefont {Dorrer}}, \bibinfo {author} {\bibfnamefont
  {B.}~\bibnamefont {Kruschwitz}}, \bibinfo {author} {\bibfnamefont
  {L.}~\bibnamefont {Waxer}}, \bibinfo {author} {\bibfnamefont {G.~J.}\
  \bibnamefont {Williams}}, \emph {et~al.},\ }\bibfield  {title} {\bibinfo
  {title} {{Microcoulomb ($0.7 \, {\pm} \, \frac{0.4}{0.2} \, \mathrm{\upmu
  C}$) laser plasma accelerator on OMEGA EP}},\ }\href
  {https://doi.org/10.1038/s41598-021-86523-5} {\bibfield  {journal} {\bibinfo
  {journal} {Sci. Rep.}\ }\textbf {\bibinfo {volume} {11}},\ \bibinfo {pages}
  {7498} (\bibinfo {year} {2021})}\BibitemShut {NoStop}%
\bibitem [{\citenamefont {Rosmej}\ \emph {et~al.}(2020)\citenamefont {Rosmej},
  \citenamefont {Gyrdymov}, \citenamefont {Günther}, \citenamefont {Andreev},
  \citenamefont {Tavana}, \citenamefont {Neumayer}, \citenamefont {Zähter},
  \citenamefont {Zahn}, \citenamefont {Popov}, \citenamefont {Borisenko} \emph
  {et~al.}}]{2020_RosmejON}%
  \BibitemOpen
  \bibfield  {author} {\bibinfo {author} {\bibfnamefont {O.~N.}\ \bibnamefont
  {Rosmej}}, \bibinfo {author} {\bibfnamefont {M.}~\bibnamefont {Gyrdymov}},
  \bibinfo {author} {\bibfnamefont {M.~M.}\ \bibnamefont {Günther}}, \bibinfo
  {author} {\bibfnamefont {N.~E.}\ \bibnamefont {Andreev}}, \bibinfo {author}
  {\bibfnamefont {P.}~\bibnamefont {Tavana}}, \bibinfo {author} {\bibfnamefont
  {P.}~\bibnamefont {Neumayer}}, \bibinfo {author} {\bibfnamefont
  {S.}~\bibnamefont {Zähter}}, \bibinfo {author} {\bibfnamefont
  {N.}~\bibnamefont {Zahn}}, \bibinfo {author} {\bibfnamefont {V.~S.}\
  \bibnamefont {Popov}}, \bibinfo {author} {\bibfnamefont {N.~G.}\ \bibnamefont
  {Borisenko}}, \emph {et~al.},\ }\bibfield  {title} {\bibinfo {title}
  {{High-current laser-driven beams of relativistic electrons for high energy
  density research}},\ }\href {https://doi.org/10.1088/1361-6587/abb24e}
  {\bibfield  {journal} {\bibinfo  {journal} {Plasma Phys. Control. Fusion}\
  }\textbf {\bibinfo {volume} {62}},\ \bibinfo {pages} {115024} (\bibinfo
  {year} {2020})}\BibitemShut {NoStop}%
\bibitem [{\citenamefont {Babjak}\ \emph
  {et~al.}(2024{\natexlab{a}})\citenamefont {Babjak}, \citenamefont
  {Willingale}, \citenamefont {Arefiev},\ and\ \citenamefont
  {Vranic}}]{2024_BabjakR}%
  \BibitemOpen
  \bibfield  {author} {\bibinfo {author} {\bibfnamefont {R.}~\bibnamefont
  {Babjak}}, \bibinfo {author} {\bibfnamefont {L.}~\bibnamefont {Willingale}},
  \bibinfo {author} {\bibfnamefont {A.}~\bibnamefont {Arefiev}},\ and\ \bibinfo
  {author} {\bibfnamefont {M.}~\bibnamefont {Vranic}},\ }\bibfield  {title}
  {\bibinfo {title} {{Direct Laser Acceleration in Underdense Plasmas with
  Multi-PW Lasers: A Path to High-Charge, GeV-Class Electron Bunches}},\ }\href
  {https://doi.org/10.1103/PhysRevLett.132.125001} {\bibfield  {journal}
  {\bibinfo  {journal} {Phys. Rev. Lett.}\ }\textbf {\bibinfo {volume} {132}},\
  \bibinfo {pages} {125001} (\bibinfo {year} {2024}{\natexlab{a}})}\BibitemShut
  {NoStop}%
\bibitem [{\citenamefont {Babjak}\ \emph
  {et~al.}(2024{\natexlab{b}})\citenamefont {Babjak}, \citenamefont {Martinez},
  \citenamefont {Krus},\ and\ \citenamefont {Vranic}}]{2024_BabjakRb}%
  \BibitemOpen
  \bibfield  {author} {\bibinfo {author} {\bibfnamefont {R.}~\bibnamefont
  {Babjak}}, \bibinfo {author} {\bibfnamefont {B.}~\bibnamefont {Martinez}},
  \bibinfo {author} {\bibfnamefont {M.}~\bibnamefont {Krus}},\ and\ \bibinfo
  {author} {\bibfnamefont {M.}~\bibnamefont {Vranic}},\ }\bibfield  {title}
  {\bibinfo {title} {{Direct laser acceleration in varying plasma density
  profiles}},\ }\href {https://doi.org/10.1088/1367-2630/ad7280} {\bibfield
  {journal} {\bibinfo  {journal} {New J. Phys.}\ }\textbf {\bibinfo {volume}
  {26}},\ \bibinfo {pages} {093002} (\bibinfo {year}
  {2024}{\natexlab{b}})}\BibitemShut {NoStop}%
\bibitem [{\citenamefont {Cohen}\ \emph {et~al.}(2024)\citenamefont {Cohen},
  \citenamefont {T.}, \citenamefont {Tangtartharakul}, \citenamefont
  {Perelmutter}, \citenamefont {Elkind}, \citenamefont {Gershuni},
  \citenamefont {Levanon}, \citenamefont {Arefiev},\ and\ \citenamefont
  {Pomerantz}}]{2024_CohenI}%
  \BibitemOpen
  \bibfield  {author} {\bibinfo {author} {\bibfnamefont {I.}~\bibnamefont
  {Cohen}}, \bibinfo {author} {\bibfnamefont {M.}~\bibnamefont {T.}}, \bibinfo
  {author} {\bibfnamefont {K.}~\bibnamefont {Tangtartharakul}}, \bibinfo
  {author} {\bibfnamefont {L.}~\bibnamefont {Perelmutter}}, \bibinfo {author}
  {\bibfnamefont {M.}~\bibnamefont {Elkind}}, \bibinfo {author} {\bibfnamefont
  {Y.}~\bibnamefont {Gershuni}}, \bibinfo {author} {\bibfnamefont
  {A.}~\bibnamefont {Levanon}}, \bibinfo {author} {\bibfnamefont {A.~V.}\
  \bibnamefont {Arefiev}},\ and\ \bibinfo {author} {\bibfnamefont
  {I.}~\bibnamefont {Pomerantz}},\ }\bibfield  {title} {\bibinfo {title}
  {{Undepleted direct laser acceleration}},\ }\href
  {https://doi.org/10.1126/sciadv.adk1947} {\bibfield  {journal} {\bibinfo
  {journal} {Sci. Adv.}\ }\textbf {\bibinfo {volume} {10}},\ \bibinfo {pages}
  {eadk1947} (\bibinfo {year} {2024})}\BibitemShut {NoStop}%
\bibitem [{\citenamefont {Tang}\ \emph {et~al.}(2024)\citenamefont {Tang},
  \citenamefont {Tangtartharakul}, \citenamefont {Babjak}, \citenamefont {Yeh},
  \citenamefont {Albert}, \citenamefont {Chen}, \citenamefont {Campbell},
  \citenamefont {Ma}, \citenamefont {Nilson}, \citenamefont {Russell} \emph
  {et~al.}}]{2024_TangH}%
  \BibitemOpen
  \bibfield  {author} {\bibinfo {author} {\bibfnamefont {H.}~\bibnamefont
  {Tang}}, \bibinfo {author} {\bibfnamefont {K.}~\bibnamefont
  {Tangtartharakul}}, \bibinfo {author} {\bibfnamefont {R.}~\bibnamefont
  {Babjak}}, \bibinfo {author} {\bibfnamefont {I.~L.}\ \bibnamefont {Yeh}},
  \bibinfo {author} {\bibfnamefont {F.}~\bibnamefont {Albert}}, \bibinfo
  {author} {\bibfnamefont {H.}~\bibnamefont {Chen}}, \bibinfo {author}
  {\bibfnamefont {P.~T.}\ \bibnamefont {Campbell}}, \bibinfo {author}
  {\bibfnamefont {Y.}~\bibnamefont {Ma}}, \bibinfo {author} {\bibfnamefont
  {P.~M.}\ \bibnamefont {Nilson}}, \bibinfo {author} {\bibfnamefont {B.~K.}\
  \bibnamefont {Russell}}, \emph {et~al.},\ }\bibfield  {title} {\bibinfo
  {title} {{The influence of laser focusing conditions on the direct laser
  acceleration of electrons}},\ }\href
  {https://doi.org/10.1088/1367-2630/ad3be4} {\bibfield  {journal} {\bibinfo
  {journal} {New J. Phys.}\ }\textbf {\bibinfo {volume} {26}},\ \bibinfo
  {pages} {053010} (\bibinfo {year} {2024})}\BibitemShut {NoStop}%
\bibitem [{\citenamefont {Rosmej}\ \emph {et~al.}(2025)\citenamefont {Rosmej},
  \citenamefont {Gyrdymov}, \citenamefont {Andreev}, \citenamefont {Tavana},
  \citenamefont {Popov}, \citenamefont {Borisenko}, \citenamefont {Gromov},
  \citenamefont {Gus’kov}, \citenamefont {Yakhin}, \citenamefont {Vegunova}
  \emph {et~al.}}]{2025_RosmejON}%
  \BibitemOpen
  \bibfield  {author} {\bibinfo {author} {\bibfnamefont {O.~N.}\ \bibnamefont
  {Rosmej}}, \bibinfo {author} {\bibfnamefont {M.}~\bibnamefont {Gyrdymov}},
  \bibinfo {author} {\bibfnamefont {N.~E.}\ \bibnamefont {Andreev}}, \bibinfo
  {author} {\bibfnamefont {P.}~\bibnamefont {Tavana}}, \bibinfo {author}
  {\bibfnamefont {V.}~\bibnamefont {Popov}}, \bibinfo {author} {\bibfnamefont
  {N.~G.}\ \bibnamefont {Borisenko}}, \bibinfo {author} {\bibfnamefont {A.~I.}\
  \bibnamefont {Gromov}}, \bibinfo {author} {\bibfnamefont {S.~Y.}\
  \bibnamefont {Gus’kov}}, \bibinfo {author} {\bibfnamefont {R.}~\bibnamefont
  {Yakhin}}, \bibinfo {author} {\bibfnamefont {G.~A.}\ \bibnamefont
  {Vegunova}}, \emph {et~al.},\ }\bibfield  {title} {\bibinfo {title}
  {{Advanced plasma target from pre-ionized low-density foam for effective and
  robust direct laser acceleration of electrons}},\ }\href
  {https://doi.org/10.1017/hpl.2024.85} {\bibfield  {journal} {\bibinfo
  {journal} {High Power Laser Sci. Eng.}\ }\textbf {\bibinfo {volume} {13}},\
  \bibinfo {pages} {e3} (\bibinfo {year} {2025})}\BibitemShut {NoStop}%
\bibitem [{\citenamefont {Stark}\ \emph {et~al.}(2016)\citenamefont {Stark},
  \citenamefont {Toncian},\ and\ \citenamefont {Arefiev}}]{2016_StarkDJ}%
  \BibitemOpen
  \bibfield  {author} {\bibinfo {author} {\bibfnamefont {D.~J.}\ \bibnamefont
  {Stark}}, \bibinfo {author} {\bibfnamefont {T.}~\bibnamefont {Toncian}},\
  and\ \bibinfo {author} {\bibfnamefont {A.~V.}\ \bibnamefont {Arefiev}},\
  }\bibfield  {title} {\bibinfo {title} {{Enhanced Multi-MeV Photon Emission by
  a Laser-Driven Electron Beam in a Self-Generated Magnetic Field}},\ }\href
  {https://doi.org/10.1103/PhysRevLett.116.185003} {\bibfield  {journal}
  {\bibinfo  {journal} {Phys. Rev. Lett.}\ }\textbf {\bibinfo {volume} {116}},\
  \bibinfo {pages} {185003} (\bibinfo {year} {2016})}\BibitemShut {NoStop}%
\bibitem [{\citenamefont {Vranic}\ \emph {et~al.}(2018)\citenamefont {Vranic},
  \citenamefont {Fonseca},\ and\ \citenamefont {Silva}}]{2018_VranicM}%
  \BibitemOpen
  \bibfield  {author} {\bibinfo {author} {\bibfnamefont {M.}~\bibnamefont
  {Vranic}}, \bibinfo {author} {\bibfnamefont {R.~A.}\ \bibnamefont
  {Fonseca}},\ and\ \bibinfo {author} {\bibfnamefont {L.~O.}\ \bibnamefont
  {Silva}},\ }\bibfield  {title} {\bibinfo {title} {{Extremely intense
  laser-based electron acceleration in a plasma channel}},\ }\href
  {https://doi.org/10.1088/1361-6587/aaa36c} {\bibfield  {journal} {\bibinfo
  {journal} {Plasma Phys. Control. Fusion}\ }\textbf {\bibinfo {volume} {60}},\
  \bibinfo {pages} {034002} (\bibinfo {year} {2018})}\BibitemShut {NoStop}%
\bibitem [{\citenamefont {Gong}\ \emph {et~al.}(2020)\citenamefont {Gong},
  \citenamefont {Mackenroth}, \citenamefont {Wang}, \citenamefont {Yan},
  \citenamefont {Toncian},\ and\ \citenamefont {Arefiev}}]{2020_GongZ}%
  \BibitemOpen
  \bibfield  {author} {\bibinfo {author} {\bibfnamefont {Z.}~\bibnamefont
  {Gong}}, \bibinfo {author} {\bibfnamefont {F.}~\bibnamefont {Mackenroth}},
  \bibinfo {author} {\bibfnamefont {T.}~\bibnamefont {Wang}}, \bibinfo {author}
  {\bibfnamefont {X.~Q.}\ \bibnamefont {Yan}}, \bibinfo {author} {\bibfnamefont
  {T.}~\bibnamefont {Toncian}},\ and\ \bibinfo {author} {\bibfnamefont {A.~V.}\
  \bibnamefont {Arefiev}},\ }\bibfield  {title} {\bibinfo {title} {Direct laser
  acceleration of electrons assisted by strong laser-driven azimuthal plasma
  magnetic fields},\ }\href {https://doi.org/10.1103/PhysRevE.102.013206}
  {\bibfield  {journal} {\bibinfo  {journal} {Phys. Rev. E}\ }\textbf {\bibinfo
  {volume} {102}},\ \bibinfo {pages} {013206} (\bibinfo {year}
  {2020})}\BibitemShut {NoStop}%
\bibitem [{\citenamefont {Jirka}\ \emph {et~al.}(2020)\citenamefont {Jirka},
  \citenamefont {Vranic}, \citenamefont {Grismayer},\ and\ \citenamefont
  {Silva}}]{2020_JirkaM}%
  \BibitemOpen
  \bibfield  {author} {\bibinfo {author} {\bibfnamefont {M.}~\bibnamefont
  {Jirka}}, \bibinfo {author} {\bibfnamefont {M.}~\bibnamefont {Vranic}},
  \bibinfo {author} {\bibfnamefont {T.}~\bibnamefont {Grismayer}},\ and\
  \bibinfo {author} {\bibfnamefont {L.~O.}\ \bibnamefont {Silva}},\ }\bibfield
  {title} {\bibinfo {title} {{Scaling laws for direct laser acceleration in a
  radiation-reaction dominated regime}},\ }\href
  {https://doi.org/10.1088/1367-2630/aba653} {\bibfield  {journal} {\bibinfo
  {journal} {New J. Phys.}\ }\textbf {\bibinfo {volume} {22}},\ \bibinfo
  {pages} {083058} (\bibinfo {year} {2020})}\BibitemShut {NoStop}%
\bibitem [{\citenamefont {Martinez}\ \emph {et~al.}(2023)\citenamefont
  {Martinez}, \citenamefont {Barbosa},\ and\ \citenamefont
  {Vranic}}]{2023_MartinezB}%
  \BibitemOpen
  \bibfield  {author} {\bibinfo {author} {\bibfnamefont {B.}~\bibnamefont
  {Martinez}}, \bibinfo {author} {\bibfnamefont {B.}~\bibnamefont {Barbosa}},\
  and\ \bibinfo {author} {\bibfnamefont {M.}~\bibnamefont {Vranic}},\
  }\bibfield  {title} {\bibinfo {title} {{Creation and direct laser
  acceleration of positrons in a single stage}},\ }\href
  {https://doi.org/10.1103/PhysRevAccelBeams.26.011301} {\bibfield  {journal}
  {\bibinfo  {journal} {Phys. Rev. Accel. Beams}\ }\textbf {\bibinfo {volume}
  {26}},\ \bibinfo {pages} {011301} (\bibinfo {year} {2023})}\BibitemShut
  {NoStop}%
\bibitem [{\citenamefont {Valenta}\ \emph {et~al.}(2024)\citenamefont
  {Valenta}, \citenamefont {Maslarova}, \citenamefont {Babjak}, \citenamefont
  {Martinez}, \citenamefont {Bulanov},\ and\ \citenamefont
  {Vrani\ifmmode~\acute{c}\else \'{c}\fi{}}}]{2024_ValentaP}%
  \BibitemOpen
  \bibfield  {author} {\bibinfo {author} {\bibfnamefont {P.}~\bibnamefont
  {Valenta}}, \bibinfo {author} {\bibfnamefont {D.}~\bibnamefont {Maslarova}},
  \bibinfo {author} {\bibfnamefont {R.}~\bibnamefont {Babjak}}, \bibinfo
  {author} {\bibfnamefont {B.}~\bibnamefont {Martinez}}, \bibinfo {author}
  {\bibfnamefont {S.~V.}\ \bibnamefont {Bulanov}},\ and\ \bibinfo {author}
  {\bibfnamefont {M.}~\bibnamefont {Vrani\ifmmode~\acute{c}\else \'{c}\fi{}}},\
  }\bibfield  {title} {\bibinfo {title} {{Direct laser acceleration: A model
  for the electron injection from the walls of a cylindrical guiding
  structure}},\ }\href {https://doi.org/10.1103/PhysRevE.109.065204} {\bibfield
   {journal} {\bibinfo  {journal} {Phys. Rev. E}\ }\textbf {\bibinfo {volume}
  {109}},\ \bibinfo {pages} {065204} (\bibinfo {year} {2024})}\BibitemShut
  {NoStop}%
\bibitem [{\citenamefont {Arefiev}\ \emph {et~al.}(2016)\citenamefont
  {Arefiev}, \citenamefont {Khudik}, \citenamefont {Robinson}, \citenamefont
  {Shvets}, \citenamefont {Willingale},\ and\ \citenamefont
  {Schollmeier}}]{2016_ArefievAV}%
  \BibitemOpen
  \bibfield  {author} {\bibinfo {author} {\bibfnamefont {A.~V.}\ \bibnamefont
  {Arefiev}}, \bibinfo {author} {\bibfnamefont {V.~N.}\ \bibnamefont {Khudik}},
  \bibinfo {author} {\bibfnamefont {A.~P.~L.}\ \bibnamefont {Robinson}},
  \bibinfo {author} {\bibfnamefont {G.}~\bibnamefont {Shvets}}, \bibinfo
  {author} {\bibfnamefont {L.}~\bibnamefont {Willingale}},\ and\ \bibinfo
  {author} {\bibfnamefont {M.}~\bibnamefont {Schollmeier}},\ }\bibfield
  {title} {\bibinfo {title} {{Beyond the ponderomotive limit: Direct laser
  acceleration of relativistic electrons in sub-critical plasmas}},\ }\href
  {https://doi.org/10.1063/1.4946024} {\bibfield  {journal} {\bibinfo
  {journal} {Phys. Plasmas}\ }\textbf {\bibinfo {volume} {23}},\ \bibinfo
  {pages} {056704} (\bibinfo {year} {2016})}\BibitemShut {NoStop}%
\bibitem [{\citenamefont {Li}\ \emph {et~al.}(2021)\citenamefont {Li},
  \citenamefont {Singh}, \citenamefont {Palaniyappan},\ and\ \citenamefont
  {Huang}}]{2021_LiFY}%
  \BibitemOpen
  \bibfield  {author} {\bibinfo {author} {\bibfnamefont {F.~Y.}\ \bibnamefont
  {Li}}, \bibinfo {author} {\bibfnamefont {P.~K.}\ \bibnamefont {Singh}},
  \bibinfo {author} {\bibfnamefont {S.}~\bibnamefont {Palaniyappan}},\ and\
  \bibinfo {author} {\bibfnamefont {C.~K.}\ \bibnamefont {Huang}},\ }\bibfield
  {title} {\bibinfo {title} {{Particle resonances and trapping of direct laser
  acceleration in a laser-plasma channel}},\ }\href
  {https://doi.org/10.1103/PhysRevAccelBeams.24.041301} {\bibfield  {journal}
  {\bibinfo  {journal} {Phys. Rev. Accel. Beams}\ }\textbf {\bibinfo {volume}
  {24}},\ \bibinfo {pages} {041301} (\bibinfo {year} {2021})}\BibitemShut
  {NoStop}%
\bibitem [{\citenamefont {Jeong}\ \emph {et~al.}(2018)\citenamefont {Jeong},
  \citenamefont {Bulanov}, \citenamefont {Weber},\ and\ \citenamefont
  {Korn}}]{2018_JeongTM}%
  \BibitemOpen
  \bibfield  {author} {\bibinfo {author} {\bibfnamefont {T.~M.}\ \bibnamefont
  {Jeong}}, \bibinfo {author} {\bibfnamefont {S.~V.}\ \bibnamefont {Bulanov}},
  \bibinfo {author} {\bibfnamefont {S.}~\bibnamefont {Weber}},\ and\ \bibinfo
  {author} {\bibfnamefont {G.}~\bibnamefont {Korn}},\ }\bibfield  {title}
  {\bibinfo {title} {{Analysis on the longitudinal field strength formed by
  tightly-focused radially-polarized femtosecond petawatt laser pulse}},\
  }\href {https://doi.org/10.1364/OE.26.033091} {\bibfield  {journal} {\bibinfo
   {journal} {Opt. Express}\ }\textbf {\bibinfo {volume} {26}},\ \bibinfo
  {pages} {33091} (\bibinfo {year} {2018})}\BibitemShut {NoStop}%
\bibitem [{\citenamefont {Pozar}(2011)}]{2011_PozarM}%
  \BibitemOpen
  \bibfield  {author} {\bibinfo {author} {\bibfnamefont {D.~M.}\ \bibnamefont
  {Pozar}},\ }\href@noop {} {\emph {\bibinfo {title} {Microwave
  Engineering}}},\ \bibinfo {edition} {4th}\ ed.\ (\bibinfo  {publisher} {John
  Wiley \& Sons},\ \bibinfo {address} {Hoboken, NJ},\ \bibinfo {year}
  {2011})\BibitemShut {NoStop}%
\bibitem [{\citenamefont {Kapchinskiy}(1984)}]{1984_KapchinskiyIM}%
  \BibitemOpen
  \bibfield  {author} {\bibinfo {author} {\bibfnamefont {I.~M.}\ \bibnamefont
  {Kapchinskiy}},\ }\href {https://www.osti.gov/biblio/5520246} {\emph
  {\bibinfo {title} {{Theory of Linear Resonance Accelerators}}}}\ (\bibinfo
  {publisher} {Harwood Academic Publishers},\ \bibinfo {address} {New York,
  NY},\ \bibinfo {year} {1984})\BibitemShut {NoStop}%
\bibitem [{\citenamefont {Humphries}(1999)}]{1999_HumphriesS}%
  \BibitemOpen
  \bibfield  {author} {\bibinfo {author} {\bibfnamefont {S.}~\bibnamefont
  {Humphries}},\ }\href@noop {} {\emph {\bibinfo {title} {{Principles of
  Charged Particle Acceleration}}}}\ (\bibinfo  {publisher} {Wiley},\ \bibinfo
  {address} {New York},\ \bibinfo {year} {1999})\BibitemShut {NoStop}%
\bibitem [{\citenamefont {Arber}\ \emph {et~al.}(2015)\citenamefont {Arber},
  \citenamefont {Bennett}, \citenamefont {Brady}, \citenamefont
  {Lawrence-Douglas}, \citenamefont {Ramsay}, \citenamefont {Sircombe},
  \citenamefont {Gillies}, \citenamefont {Evans}, \citenamefont {Schmitz},
  \citenamefont {Bell} \emph {et~al.}}]{2015_ArberTD}%
  \BibitemOpen
  \bibfield  {author} {\bibinfo {author} {\bibfnamefont {T.~D.}\ \bibnamefont
  {Arber}}, \bibinfo {author} {\bibfnamefont {K.}~\bibnamefont {Bennett}},
  \bibinfo {author} {\bibfnamefont {C.~S.}\ \bibnamefont {Brady}}, \bibinfo
  {author} {\bibfnamefont {A.}~\bibnamefont {Lawrence-Douglas}}, \bibinfo
  {author} {\bibfnamefont {M.~G.}\ \bibnamefont {Ramsay}}, \bibinfo {author}
  {\bibfnamefont {N.~J.}\ \bibnamefont {Sircombe}}, \bibinfo {author}
  {\bibfnamefont {P.}~\bibnamefont {Gillies}}, \bibinfo {author} {\bibfnamefont
  {R.~G.}\ \bibnamefont {Evans}}, \bibinfo {author} {\bibfnamefont
  {H.}~\bibnamefont {Schmitz}}, \bibinfo {author} {\bibfnamefont {A.~R.}\
  \bibnamefont {Bell}}, \emph {et~al.},\ }\bibfield  {title} {\bibinfo {title}
  {{Contemporary particle-in-cell approach to laser-plasma modelling}},\ }\href
  {https://doi.org/10.1088/0741-3335/57/11/113001} {\bibfield  {journal}
  {\bibinfo  {journal} {Plasma Phys. Control. Fusion}\ }\textbf {\bibinfo
  {volume} {57}},\ \bibinfo {pages} {113001} (\bibinfo {year}
  {2015})}\BibitemShut {NoStop}%
\bibitem [{\citenamefont {Snyder}\ \emph {et~al.}(2019)\citenamefont {Snyder},
  \citenamefont {Ji}, \citenamefont {George}, \citenamefont {Willis},
  \citenamefont {Cochran}, \citenamefont {Daskalova}, \citenamefont {Handler},
  \citenamefont {Rubin}, \citenamefont {Poole}, \citenamefont {Nasir} \emph
  {et~al.}}]{2019_SnyderJ}%
  \BibitemOpen
  \bibfield  {author} {\bibinfo {author} {\bibfnamefont {J.}~\bibnamefont
  {Snyder}}, \bibinfo {author} {\bibfnamefont {L.~L.}\ \bibnamefont {Ji}},
  \bibinfo {author} {\bibfnamefont {K.~M.}\ \bibnamefont {George}}, \bibinfo
  {author} {\bibfnamefont {C.}~\bibnamefont {Willis}}, \bibinfo {author}
  {\bibfnamefont {G.~E.}\ \bibnamefont {Cochran}}, \bibinfo {author}
  {\bibfnamefont {R.~L.}\ \bibnamefont {Daskalova}}, \bibinfo {author}
  {\bibfnamefont {A.}~\bibnamefont {Handler}}, \bibinfo {author} {\bibfnamefont
  {T.}~\bibnamefont {Rubin}}, \bibinfo {author} {\bibfnamefont {P.~L.}\
  \bibnamefont {Poole}}, \bibinfo {author} {\bibfnamefont {D.}~\bibnamefont
  {Nasir}}, \emph {et~al.},\ }\bibfield  {title} {\bibinfo {title}
  {{Relativistic laser driven electron accelerator using micro-channel plasma
  targets}},\ }\href {https://doi.org/10.1063/1.5087409} {\bibfield  {journal}
  {\bibinfo  {journal} {Phys. Plasmas}\ }\textbf {\bibinfo {volume} {26}},\
  \bibinfo {pages} {033110} (\bibinfo {year} {2019})}\BibitemShut {NoStop}%
\bibitem [{\citenamefont {Wang}\ \emph {et~al.}(2021)\citenamefont {Wang},
  \citenamefont {Blackman}, \citenamefont {Chin},\ and\ \citenamefont
  {Arefiev}}]{2021_WangT}%
  \BibitemOpen
  \bibfield  {author} {\bibinfo {author} {\bibfnamefont {T.}~\bibnamefont
  {Wang}}, \bibinfo {author} {\bibfnamefont {D.}~\bibnamefont {Blackman}},
  \bibinfo {author} {\bibfnamefont {K.}~\bibnamefont {Chin}},\ and\ \bibinfo
  {author} {\bibfnamefont {A.}~\bibnamefont {Arefiev}},\ }\bibfield  {title}
  {\bibinfo {title} {{Effects of simulation dimensionality on laser-driven
  electron acceleration and photon emission in hollow microchannel targets}},\
  }\href {https://doi.org/10.1103/PhysRevE.104.045206} {\bibfield  {journal}
  {\bibinfo  {journal} {Phys. Rev. E}\ }\textbf {\bibinfo {volume} {104}},\
  \bibinfo {pages} {045206} (\bibinfo {year} {2021})}\BibitemShut {NoStop}%
\bibitem [{\citenamefont {Shaw}\ \emph {et~al.}(2017)\citenamefont {Shaw},
  \citenamefont {Lemos}, \citenamefont {Amorim}, \citenamefont
  {Vafaei-Najafabadi}, \citenamefont {Marsh}, \citenamefont {Tsung},
  \citenamefont {Mori},\ and\ \citenamefont {Joshi}}]{2017_ShawJL}%
  \BibitemOpen
  \bibfield  {author} {\bibinfo {author} {\bibfnamefont {J.~L.}\ \bibnamefont
  {Shaw}}, \bibinfo {author} {\bibfnamefont {N.}~\bibnamefont {Lemos}},
  \bibinfo {author} {\bibfnamefont {L.~D.}\ \bibnamefont {Amorim}}, \bibinfo
  {author} {\bibfnamefont {N.}~\bibnamefont {Vafaei-Najafabadi}}, \bibinfo
  {author} {\bibfnamefont {K.~A.}\ \bibnamefont {Marsh}}, \bibinfo {author}
  {\bibfnamefont {F.~S.}\ \bibnamefont {Tsung}}, \bibinfo {author}
  {\bibfnamefont {W.~B.}\ \bibnamefont {Mori}},\ and\ \bibinfo {author}
  {\bibfnamefont {C.}~\bibnamefont {Joshi}},\ }\bibfield  {title} {\bibinfo
  {title} {{Role of Direct Laser Acceleration of Electrons in a Laser Wakefield
  Accelerator with Ionization Injection}},\ }\href
  {https://doi.org/10.1103/PhysRevLett.118.064801} {\bibfield  {journal}
  {\bibinfo  {journal} {Phys. Rev. Lett.}\ }\textbf {\bibinfo {volume} {118}},\
  \bibinfo {pages} {064801} (\bibinfo {year} {2017})}\BibitemShut {NoStop}%
\bibitem [{\citenamefont {Shaw}\ \emph {et~al.}(2018)\citenamefont {Shaw},
  \citenamefont {Lemos}, \citenamefont {Marsh}, \citenamefont {Froula},\ and\
  \citenamefont {Joshi}}]{2018_ShawJL}%
  \BibitemOpen
  \bibfield  {author} {\bibinfo {author} {\bibfnamefont {J.~L.}\ \bibnamefont
  {Shaw}}, \bibinfo {author} {\bibfnamefont {N.}~\bibnamefont {Lemos}},
  \bibinfo {author} {\bibfnamefont {K.~A.}\ \bibnamefont {Marsh}}, \bibinfo
  {author} {\bibfnamefont {D.~H.}\ \bibnamefont {Froula}},\ and\ \bibinfo
  {author} {\bibfnamefont {C.}~\bibnamefont {Joshi}},\ }\bibfield  {title}
  {\bibinfo {title} {{Experimental signatures of
  direct-laser-acceleration-assisted laser wakefield acceleration}},\ }\href
  {https://doi.org/10.1088/1361-6587/aaade1} {\bibfield  {journal} {\bibinfo
  {journal} {Plasma Phys. Control. Fusion}\ }\textbf {\bibinfo {volume} {60}},\
  \bibinfo {pages} {044012} (\bibinfo {year} {2018})}\BibitemShut {NoStop}%
\bibitem [{\citenamefont {King}\ \emph {et~al.}(2021)\citenamefont {King},
  \citenamefont {Miller}, \citenamefont {Lemos}, \citenamefont {Shaw},
  \citenamefont {Kraus}, \citenamefont {Thibodeau}, \citenamefont {Hegelich},
  \citenamefont {Hinojosa}, \citenamefont {Michel}, \citenamefont {Joshi} \emph
  {et~al.}}]{2021_KingPM}%
  \BibitemOpen
  \bibfield  {author} {\bibinfo {author} {\bibfnamefont {P.~M.}\ \bibnamefont
  {King}}, \bibinfo {author} {\bibfnamefont {K.}~\bibnamefont {Miller}},
  \bibinfo {author} {\bibfnamefont {N.}~\bibnamefont {Lemos}}, \bibinfo
  {author} {\bibfnamefont {J.~L.}\ \bibnamefont {Shaw}}, \bibinfo {author}
  {\bibfnamefont {B.~F.}\ \bibnamefont {Kraus}}, \bibinfo {author}
  {\bibfnamefont {M.}~\bibnamefont {Thibodeau}}, \bibinfo {author}
  {\bibfnamefont {B.~M.}\ \bibnamefont {Hegelich}}, \bibinfo {author}
  {\bibfnamefont {J.}~\bibnamefont {Hinojosa}}, \bibinfo {author}
  {\bibfnamefont {P.}~\bibnamefont {Michel}}, \bibinfo {author} {\bibfnamefont
  {C.}~\bibnamefont {Joshi}}, \emph {et~al.},\ }\bibfield  {title} {\bibinfo
  {title} {{Predominant contribution of direct laser acceleration to
  high-energy electron spectra in a low-density self-modulated laser wakefield
  accelerator}},\ }\href {https://doi.org/10.1103/PhysRevAccelBeams.24.011302}
  {\bibfield  {journal} {\bibinfo  {journal} {Phys. Rev. Accel. Beams}\
  }\textbf {\bibinfo {volume} {24}},\ \bibinfo {pages} {011302} (\bibinfo
  {year} {2021})}\BibitemShut {NoStop}%
\bibitem [{\citenamefont {Salamin}(2006)}]{2006_SalaminYI}%
  \BibitemOpen
  \bibfield  {author} {\bibinfo {author} {\bibfnamefont {Y.~I.}\ \bibnamefont
  {Salamin}},\ }\bibfield  {title} {\bibinfo {title} {{Electron acceleration
  from rest in vacuum by an axicon Gaussian laser beam}},\ }\href
  {https://doi.org/10.1103/PhysRevA.73.043402} {\bibfield  {journal} {\bibinfo
  {journal} {Phys. Rev. A}\ }\textbf {\bibinfo {volume} {73}},\ \bibinfo
  {pages} {043402} (\bibinfo {year} {2006})}\BibitemShut {NoStop}%
\bibitem [{\citenamefont {Singh}\ and\ \citenamefont
  {Kumar}(2011)}]{2011_SinghKP}%
  \BibitemOpen
  \bibfield  {author} {\bibinfo {author} {\bibfnamefont {K.~P.}\ \bibnamefont
  {Singh}}\ and\ \bibinfo {author} {\bibfnamefont {M.}~\bibnamefont {Kumar}},\
  }\bibfield  {title} {\bibinfo {title} {{Electron acceleration by a radially
  polarized laser pulse during ionization of low density gases}},\ }\href
  {https://doi.org/10.1103/PhysRevSTAB.14.030401} {\bibfield  {journal}
  {\bibinfo  {journal} {Phys. Rev. ST Accel. Beams}\ }\textbf {\bibinfo
  {volume} {14}},\ \bibinfo {pages} {030401} (\bibinfo {year}
  {2011})}\BibitemShut {NoStop}%
\bibitem [{\citenamefont {Carbajo}\ \emph {et~al.}(2016)\citenamefont
  {Carbajo}, \citenamefont {Nanni}, \citenamefont {Wong}, \citenamefont
  {Moriena}, \citenamefont {Keathley}, \citenamefont {Laurent}, \citenamefont
  {Miller},\ and\ \citenamefont {K\"artner}}]{2016_CarbajoS}%
  \BibitemOpen
  \bibfield  {author} {\bibinfo {author} {\bibfnamefont {S.}~\bibnamefont
  {Carbajo}}, \bibinfo {author} {\bibfnamefont {E.~A.}\ \bibnamefont {Nanni}},
  \bibinfo {author} {\bibfnamefont {L.~J.}\ \bibnamefont {Wong}}, \bibinfo
  {author} {\bibfnamefont {G.}~\bibnamefont {Moriena}}, \bibinfo {author}
  {\bibfnamefont {P.~D.}\ \bibnamefont {Keathley}}, \bibinfo {author}
  {\bibfnamefont {G.}~\bibnamefont {Laurent}}, \bibinfo {author} {\bibfnamefont
  {R.~J.~D.}\ \bibnamefont {Miller}},\ and\ \bibinfo {author} {\bibfnamefont
  {F.~X.}\ \bibnamefont {K\"artner}},\ }\bibfield  {title} {\bibinfo {title}
  {{Direct longitudinal laser acceleration of electrons in free space}},\
  }\href {https://doi.org/10.1103/PhysRevAccelBeams.19.021303} {\bibfield
  {journal} {\bibinfo  {journal} {Phys. Rev. Accel. Beams}\ }\textbf {\bibinfo
  {volume} {19}},\ \bibinfo {pages} {021303} (\bibinfo {year}
  {2016})}\BibitemShut {NoStop}%
\bibitem [{\citenamefont {Alfv\'en}(1939)}]{1939_AlfvenH}%
  \BibitemOpen
  \bibfield  {author} {\bibinfo {author} {\bibfnamefont {H.}~\bibnamefont
  {Alfv\'en}},\ }\bibfield  {title} {\bibinfo {title} {{On the Motion of Cosmic
  Rays in Interstellar Space}},\ }\href
  {https://doi.org/10.1103/PhysRev.55.425} {\bibfield  {journal} {\bibinfo
  {journal} {Phys. Rev.}\ }\textbf {\bibinfo {volume} {55}},\ \bibinfo {pages}
  {425} (\bibinfo {year} {1939})}\BibitemShut {NoStop}%
\bibitem [{\citenamefont {Sinars}\ \emph {et~al.}(2020)\citenamefont {Sinars},
  \citenamefont {Sweeney}, \citenamefont {Alexander}, \citenamefont
  {Ampleford}, \citenamefont {Ao}, \citenamefont {Apruzese}, \citenamefont
  {Aragon}, \citenamefont {Armstrong}, \citenamefont {Austin}, \citenamefont
  {Awe} \emph {et~al.}}]{2020_SinarsDB}%
  \BibitemOpen
  \bibfield  {author} {\bibinfo {author} {\bibfnamefont {D.~B.}\ \bibnamefont
  {Sinars}}, \bibinfo {author} {\bibfnamefont {M.~A.}\ \bibnamefont {Sweeney}},
  \bibinfo {author} {\bibfnamefont {C.~S.}\ \bibnamefont {Alexander}}, \bibinfo
  {author} {\bibfnamefont {D.~J.}\ \bibnamefont {Ampleford}}, \bibinfo {author}
  {\bibfnamefont {T.}~\bibnamefont {Ao}}, \bibinfo {author} {\bibfnamefont
  {J.~P.}\ \bibnamefont {Apruzese}}, \bibinfo {author} {\bibfnamefont
  {C.}~\bibnamefont {Aragon}}, \bibinfo {author} {\bibfnamefont {D.~J.}\
  \bibnamefont {Armstrong}}, \bibinfo {author} {\bibfnamefont {K.~N.}\
  \bibnamefont {Austin}}, \bibinfo {author} {\bibfnamefont {T.~J.}\
  \bibnamefont {Awe}}, \emph {et~al.},\ }\bibfield  {title} {\bibinfo {title}
  {{Review of pulsed power-driven high energy density physics research on Z at
  Sandia}},\ }\href {https://doi.org/10.1063/5.0007476} {\bibfield  {journal}
  {\bibinfo  {journal} {Phys. Plasmas}\ }\textbf {\bibinfo {volume} {27}},\
  \bibinfo {pages} {070501} (\bibinfo {year} {2020})}\BibitemShut {NoStop}%
\bibitem [{\citenamefont {Nakamura}\ \emph {et~al.}(2012)\citenamefont
  {Nakamura}, \citenamefont {Koga}, \citenamefont {Esirkepov}, \citenamefont
  {Kando}, \citenamefont {Korn},\ and\ \citenamefont
  {Bulanov}}]{2012_NakamuraT}%
  \BibitemOpen
  \bibfield  {author} {\bibinfo {author} {\bibfnamefont {T.}~\bibnamefont
  {Nakamura}}, \bibinfo {author} {\bibfnamefont {J.~K.}\ \bibnamefont {Koga}},
  \bibinfo {author} {\bibfnamefont {T.~Z.}\ \bibnamefont {Esirkepov}}, \bibinfo
  {author} {\bibfnamefont {M.}~\bibnamefont {Kando}}, \bibinfo {author}
  {\bibfnamefont {G.}~\bibnamefont {Korn}},\ and\ \bibinfo {author}
  {\bibfnamefont {S.~V.}\ \bibnamefont {Bulanov}},\ }\bibfield  {title}
  {\bibinfo {title} {{High-Power $\ensuremath{\gamma}$-Ray Flash Generation in
  Ultraintense Laser-Plasma Interactions}},\ }\href
  {https://doi.org/10.1103/PhysRevLett.108.195001} {\bibfield  {journal}
  {\bibinfo  {journal} {Phys. Rev. Lett.}\ }\textbf {\bibinfo {volume} {108}},\
  \bibinfo {pages} {195001} (\bibinfo {year} {2012})}\BibitemShut {NoStop}%
\bibitem [{\citenamefont {Ridgers}\ \emph {et~al.}(2012)\citenamefont
  {Ridgers}, \citenamefont {Brady}, \citenamefont {Duclous}, \citenamefont
  {Kirk}, \citenamefont {Bennett}, \citenamefont {Arber}, \citenamefont
  {Robinson},\ and\ \citenamefont {Bell}}]{2012_RidgersCP}%
  \BibitemOpen
  \bibfield  {author} {\bibinfo {author} {\bibfnamefont {C.~P.}\ \bibnamefont
  {Ridgers}}, \bibinfo {author} {\bibfnamefont {C.~S.}\ \bibnamefont {Brady}},
  \bibinfo {author} {\bibfnamefont {R.}~\bibnamefont {Duclous}}, \bibinfo
  {author} {\bibfnamefont {J.~G.}\ \bibnamefont {Kirk}}, \bibinfo {author}
  {\bibfnamefont {K.}~\bibnamefont {Bennett}}, \bibinfo {author} {\bibfnamefont
  {T.~D.}\ \bibnamefont {Arber}}, \bibinfo {author} {\bibfnamefont {A.~P.~L.}\
  \bibnamefont {Robinson}},\ and\ \bibinfo {author} {\bibfnamefont {A.~R.}\
  \bibnamefont {Bell}},\ }\bibfield  {title} {\bibinfo {title} {{Dense
  Electron-Positron Plasmas and Ultraintense $\ensuremath{\gamma}$ rays from
  Laser-Irradiated Solids}},\ }\href
  {https://doi.org/10.1103/PhysRevLett.108.165006} {\bibfield  {journal}
  {\bibinfo  {journal} {Phys. Rev. Lett.}\ }\textbf {\bibinfo {volume} {108}},\
  \bibinfo {pages} {165006} (\bibinfo {year} {2012})}\BibitemShut {NoStop}%
\bibitem [{\citenamefont {Hadjisolomou}\ \emph {et~al.}(2021)\citenamefont
  {Hadjisolomou}, \citenamefont {Jeong}, \citenamefont {Valenta}, \citenamefont
  {Korn},\ and\ \citenamefont {Bulanov}}]{2021_HadjisolomouP}%
  \BibitemOpen
  \bibfield  {author} {\bibinfo {author} {\bibfnamefont {P.}~\bibnamefont
  {Hadjisolomou}}, \bibinfo {author} {\bibfnamefont {T.~M.}\ \bibnamefont
  {Jeong}}, \bibinfo {author} {\bibfnamefont {P.}~\bibnamefont {Valenta}},
  \bibinfo {author} {\bibfnamefont {G.}~\bibnamefont {Korn}},\ and\ \bibinfo
  {author} {\bibfnamefont {S.~V.}\ \bibnamefont {Bulanov}},\ }\bibfield
  {title} {\bibinfo {title} {{Gamma-ray flash generation in irradiating a thin
  foil target by a single-cycle tightly focused extreme power laser pulse}},\
  }\href {https://doi.org/10.1103/PhysRevE.104.015203} {\bibfield  {journal}
  {\bibinfo  {journal} {Phys. Rev. E}\ }\textbf {\bibinfo {volume} {104}},\
  \bibinfo {pages} {015203} (\bibinfo {year} {2021})}\BibitemShut {NoStop}%
\bibitem [{\citenamefont {Hadjisolomou}\ \emph {et~al.}(2022)\citenamefont
  {Hadjisolomou}, \citenamefont {Jeong}, \citenamefont {Valenta}, \citenamefont
  {Kolenaty}, \citenamefont {Versaci}, \citenamefont {Olšovcová},
  \citenamefont {Ridgers},\ and\ \citenamefont {Bulanov}}]{2022_HadjisolomouP}%
  \BibitemOpen
  \bibfield  {author} {\bibinfo {author} {\bibfnamefont {P.}~\bibnamefont
  {Hadjisolomou}}, \bibinfo {author} {\bibfnamefont {T.~M.}\ \bibnamefont
  {Jeong}}, \bibinfo {author} {\bibfnamefont {P.}~\bibnamefont {Valenta}},
  \bibinfo {author} {\bibfnamefont {D.}~\bibnamefont {Kolenaty}}, \bibinfo
  {author} {\bibfnamefont {R.}~\bibnamefont {Versaci}}, \bibinfo {author}
  {\bibfnamefont {V.}~\bibnamefont {Olšovcová}}, \bibinfo {author}
  {\bibfnamefont {C.~P.}\ \bibnamefont {Ridgers}},\ and\ \bibinfo {author}
  {\bibfnamefont {S.~V.}\ \bibnamefont {Bulanov}},\ }\bibfield  {title}
  {\bibinfo {title} {{Gamma-ray flash in the interaction of a tightly focused
  single-cycle ultra-intense laser pulse with a solid target}},\ }\href
  {https://doi.org/10.1017/S0022377821001318} {\bibfield  {journal} {\bibinfo
  {journal} {J. Plasma Phys.}\ }\textbf {\bibinfo {volume} {88}},\ \bibinfo
  {pages} {905880104} (\bibinfo {year} {2022})}\BibitemShut {NoStop}%
\bibitem [{\citenamefont {Hadjisolomou}\ \emph {et~al.}(2025)\citenamefont
  {Hadjisolomou}, \citenamefont {Jeong}, \citenamefont {Valenta}, \citenamefont
  {Macleod}, \citenamefont {Shaisultanov}, \citenamefont {Ridgers},\ and\
  \citenamefont {Bulanov}}]{2025_HadjisolomouP}%
  \BibitemOpen
  \bibfield  {author} {\bibinfo {author} {\bibfnamefont {P.}~\bibnamefont
  {Hadjisolomou}}, \bibinfo {author} {\bibfnamefont {T.~M.}\ \bibnamefont
  {Jeong}}, \bibinfo {author} {\bibfnamefont {P.}~\bibnamefont {Valenta}},
  \bibinfo {author} {\bibfnamefont {A.~J.}\ \bibnamefont {Macleod}}, \bibinfo
  {author} {\bibfnamefont {R.}~\bibnamefont {Shaisultanov}}, \bibinfo {author}
  {\bibfnamefont {C.~P.}\ \bibnamefont {Ridgers}},\ and\ \bibinfo {author}
  {\bibfnamefont {S.~V.}\ \bibnamefont {Bulanov}},\ }\bibfield  {title}
  {\bibinfo {title} {{Attosecond gamma-ray flashes and electron-positron pairs
  in dyadic laser interaction with microwire}},\ }\href
  {https://doi.org/10.1103/PhysRevE.111.025201} {\bibfield  {journal} {\bibinfo
   {journal} {Phys. Rev. E}\ }\textbf {\bibinfo {volume} {111}},\ \bibinfo
  {pages} {025201} (\bibinfo {year} {2025})}\BibitemShut {NoStop}%
\bibitem [{\citenamefont {Hadjisolomou}\ \emph {et~al.}(2023)\citenamefont
  {Hadjisolomou}, \citenamefont {Jeong}, \citenamefont {Kolenaty},
  \citenamefont {Macleod}, \citenamefont {Olšovcová}, \citenamefont
  {Versaci}, \citenamefont {Ridgers},\ and\ \citenamefont
  {Bulanov}}]{2023_HadjisolomouP}%
  \BibitemOpen
  \bibfield  {author} {\bibinfo {author} {\bibfnamefont {P.}~\bibnamefont
  {Hadjisolomou}}, \bibinfo {author} {\bibfnamefont {T.~M.}\ \bibnamefont
  {Jeong}}, \bibinfo {author} {\bibfnamefont {D.}~\bibnamefont {Kolenaty}},
  \bibinfo {author} {\bibfnamefont {A.~J.}\ \bibnamefont {Macleod}}, \bibinfo
  {author} {\bibfnamefont {V.}~\bibnamefont {Olšovcová}}, \bibinfo {author}
  {\bibfnamefont {R.}~\bibnamefont {Versaci}}, \bibinfo {author} {\bibfnamefont
  {C.~P.}\ \bibnamefont {Ridgers}},\ and\ \bibinfo {author} {\bibfnamefont
  {S.~V.}\ \bibnamefont {Bulanov}},\ }\bibfield  {title} {\bibinfo {title}
  {{Gamma-flash generation in multi-petawatt laser–matter interactions}},\
  }\href {https://doi.org/10.1063/5.0158264} {\bibfield  {journal} {\bibinfo
  {journal} {Phys. Plasmas}\ }\textbf {\bibinfo {volume} {30}},\ \bibinfo
  {pages} {093103} (\bibinfo {year} {2023})}\BibitemShut {NoStop}%
\end{thebibliography}%

\end{document}